\documentclass[aps,prb,twocolumn,groupedaddress,draft,
              showpacs,showkeys,a4paper]{revtex4}
\usepackage{amsmath}
\usepackage[final]{graphicx} 
\begin{document}


\title{A Two-dimensional Superconductor in a Tilted Magnetic
Field - new states with finite Cooper-pair momentum}



\author{U.~Klein}
\email[]{ulf.klein@jku.at}
\affiliation{Johannes Kepler Universit{\"a}t Linz, Institut 
f{\"u}r 
Theoretische Physik, A-4040 Linz, Austria}


\date{\today}

\begin{abstract}

Varying the angle $\theta$ between applied field and the  
conducting planes of a layered superconductor in a small 
interval close to the plane-parallel field direction, a 
large number of superconducting states with unusual 
properties may be produced. For these states, the 
pair breaking effect of the magnetic field affects 
both the orbital and the spin degree of freedom. This 
leads to pair wave functions with finite momentum,
which are labeled by Landau quantum numbers 
$0<n<\infty$. The stable order parameter structure 
and magnetic field distribution for these states 
is found  by minimizing the quasiclassical free energy 
near $H_{c2}$ including nonlinear terms. One finds 
states with coexisting line-like and point-like 
order parameter zeros and states with coexisting vortices 
and antivortices. The magnetic response may be  
diamagnetic or paramagnetic depending on the position 
within the unit cell. The structure of the 
Fulde-Ferrell-Larkin-Ovchinnikov (FFLO) states 
at $\theta=0$ is reconsidered. The transition 
$n\to\infty$ of the paramagnetic vortex states to 
the FFLO-limit is analyzed and the physical reason 
for the occupation of higher Landau levels is 
pointed out. 

\end{abstract}


\maketitle

\section{\label{sec:intro}Introduction}
In this paper a theoretical study of a two-dimensional, 
clean-limit superconductor in a tilted magnetic field is 
presented. Such systems exist in nature; several
classes of layered superconductors of high purity 
with conducting planes of atomic thickness and nearly 
perfect decoupling of adjacent planes have been 
investigated in recent years. These include, among many 
others, the intercalated transition metal dichalcogenide 
$TaS_2-(pyridine)$, the organic superconductor 
$\kappa-(BEDT-TTF)_2Cu(NCS)_2$, and the magnetic field 
induced 
superconductor $\lambda-(BETS)_2FeCl_4$.

Depending on the angle $\theta$ between applied field 
and conducting planes the nature of the pair-breaking
mechanism limiting the superconducting state can be 
continuously varied. For large $\theta$ the usual orbital
pair-breaking mechanism dominates and the equilibrium state 
is the ordinary vortex lattice. With decreasing $\theta$,
in a small interval close to the parallel direction, 
spin pair-breaking becomes of a magnitude comparable to the 
orbital effect and both mechanisms must be taken into account. 
For the plane-parallel field direction, $\theta=0$, the orbital 
effect vanishes completely and the superconducting 
state is solely limited by paramagnetic pair breaking. 
The superconducting state expected in this limit 
is the Fulde-Ferrell-Larkin-Ovchinnikov (FFLO) 
state\cite{FUFE,LAOV}. The tilted-field arrangement, which 
allows to control externally the relative strength of 
both pair-breaking mechanisms, has first been investigated 
by Bulaevskii\cite{BULATILT}. 

The upper critical field $H_{c2}$, where a second order phase
transition between the normal-conducting and the 
superconducting state takes place, has been calculated 
for arbitrary angle $\theta$ and temperature $T=0$ by 
Bulaevskii\cite{BULATILT}. This treatment was generalized 
to arbitrary $T$ by Shimahara and 
Rainer \cite{SHIMARAIN} . The field $H_{c2}$ has 
a cusp-like shape, considered both as a function of $\theta$ or 
$T$, with different pieces of the curve belonging to 
different values of the Landau quantum number 
$n$ ($n=0,1,\ldots$). In the orbital pair breaking regime, for 
large $\theta$, one finds as expected $n=0$. As is well known, 
this lowest value $n=0$ determines the (orbital) upper 
critical field of the familiar vortex state, both in 
the framework of Ginzburg-Landau(GL)- and microscopic 
theories of superconductivity.  
With decreasing $\theta$, higher-$n$ segments of the critical 
field curve appear close to the plane-parallel 
orientation. For  $\theta\to0$ one finds\cite{SHIMARAIN} 
$n\to\infty$ and agreement with the FFLO upper critical field. 
Thus, in this purely paramagnetic limit, the stable 
state below $H_{c2}$ must be the FFLO state.

Paramagnetically-limited superconductivity
differs in fundamental aspects, such as Meissner effect 
and spin-polarization, from the behavior of 
the usual, orbitally-limited superconducting state.
In the FFLO state pairing takes place between electrons 
with momentum and spin values $(\vec{k}+\vec{q}/2,\uparrow)$ 
and $(-\vec{k}+\vec{q}/2,\downarrow)$. This leads to Cooper pairs 
with finite momentum $\hbar\vec{q}$ and a spatially 
inhomogeneous superconducting order parameter given by 
$\Delta(\vec{r})=\Delta_0\exp(\imath \vec{q}\vec{r})$ (or by linear 
combinations of such terms with the same absolute 
value of $\vec{q}$). The pair-breaking is entirely  
due to the Zeeman coupling between the magnetic moment 
$\mu$ of the electrons  and the external magnetic field 
$\vec{H}$. The general rule, for bulk superconducting 
states, that gradient terms in the free energy must 
only be taken into account if a nontrivial vector 
potential is present breaks down for the FFLO state. 

At $T=0$, the Cooper pair momentum of the FFLO state is 
approximately given by $\hbar q = |p_{F\uparrow}-p_{F\downarrow }| $, where 
$ |p_{F\uparrow}-p_{F\downarrow }|=\mu H\sqrt{2m/E_F}$ is the difference in Fermi 
momentum between spin-up and spin-down electrons. With 
increasing $T$ the FFLO wave number $q$ decreases and 
vanishes at the tricritical point $T_{tri}=0.56\,T_c$. The FFLO 
state is only stable for $T<T_{tri}$, where its upper critical 
field $H_{FFLO}$ exceeds the Pauli limiting field $H_P$ of 
the homogeneous superconducting state\cite{CHANDRA,CLOGSTON}. 
At $T=0$, $\mu H_P=\Delta_0/\sqrt{2}$, where $\Delta_0$ is the superconducting 
gap at $T=0$. The second order phase transition line 
$H_{FFLO}(T)$ depends on the shape of the Fermi surface. In 
this paper we use a cylindrical Fermi surface appropriate 
for a two-dimensional (2D) geometry. The corresponding critical 
field\cite{BURKRAIN} is given by $\mu H_{FFLO}=\Delta_0$ at $T=0$. 

Between the ordinary vortex state with $n=0$ and the FFLO
state with $n\to\infty$ a countable infinite number of 
unconventional superconducting states, characterized by 
Landau quantum numbers $n=1,2,\ldots$, exist. The transition 
from the vortex state to the first of these, the $n=1$ 
state, occurs at an angle $\theta_1$ given approximately by
\begin{equation}
  \label{eq:docan1}
\sin\theta_1 \approx \frac{H_{c2}^{orb}}{H_P}\approx \frac{k_BT_c}{mv_F^2}
\mbox{,}
\end{equation}
where $H_{c2}^{orb}$ and $H_p$ are the ''pure'' orbital and 
paramagnetic upper critical fields, respectively. Since 
$H_p \gg H_{c2}^{orb}$ the experimental upper critical field for a 
three-dimensional sample is given by $H_{c2}^{orb}$. 
Because $\theta_1\ll1$ (generally $\theta_1$ will be of the order 
of magnitude of $1$ Degree), the perpendicular component
$H_{\perp}=H\sin\theta$ for all of these states with $n>0$ will be 
much smaller than the parallel component $H_\|=H\cos\theta$. Thus, 
these states will have some properties in common with the 
FFLO state, namely strong paramagnetic-pair breaking, a 
spatially inhomogeneous order parameter, and Cooper pairs 
with finite velocity of the center of mass coordinate. 
Despite this similarity with regard to general features, 
the order parameter structure for the $n>0$ states may be 
completely different, even for large $n$, from the FFLO 
state. The reason is, that a finite perpendicular component 
$H_\perp$, no matter how small, implies a new and rather 
stringent topological constraint on the equilibrium 
structure, namely the flux quantization condition. The 
subject of the present paper is the detailed investigation 
of the structure of these $n>0$ states, which might be 
referred to either as FFLO precursor states or as 
paramagnetic vortex states, in the vicinity of the 
upper critical field $H_{c2}$. A theoretical treatment of
these FFLO precursor states, reporting several essential 
results and an outline of the calculation, has been 
published previously\cite{KLRAISHI}. This 
paper\cite{KLRAISHI} will be referred to as KRS in 
what follows. In the present paper many new results 
are reported and the treatment is extended with regard 
to several points, including finite values of $\kappa$, the 
purely paramagnetic limit $\theta=0$, and the transition $n\to\infty$. 

It should be pointed out, that the physical origin of the 
Landau level quantization effects for Cooper pairs, 
considered in the present paper, is very different 
from the  Landau quantization effects for single 
electron states discussed in a large number of 
publications by Tesanovic et al.\cite{RATEREV}, 
Rajagopal et al.\cite{RYARAJ1}, Norman et al.\cite{AKDOGINO} 
and others. The latter are mainly concerned with the 
relative-coordinate degree of freedom of the two bound 
electrons constituting a Cooper pair and lead to 
measurable consequences only outside the range of 
validity of the quasiclassical approximation, at very 
low temperature $T < (k_BT_c)^2/E_F$ and/or high fields.
In addition, a mechanism is required to 
suppress the Zeeman effect, which is neglected in the 
theoretical treatment and is not compatible 
with the predicted phenomena. The question whether 
the most dramatic consequences\cite{TERAXI}
(reentrant superconductivity) of this type of Landau 
quantization  effects will be observable, has been the 
subject of a controversial 
discussion\cite{RIESCHKLE,NICKUM}. In contrast, 
the present Landau level quantization mechanism is 
a \emph{consequence} of the Zeeman effect, concerns the 
center of mass motion of the Cooper pairs, and can 
be described (as will be discussed shortly) by means 
of the quasiclassical theory of superconductivity.

Restricting ourselves to the vicinity of the upper critical 
field $H_{c2}$ we may use an expansion of the free energy in 
powers of the order parameter $\Delta$, keeping only a finite 
number of terms. An analogous gradient-expansion, which 
would lead to a relatively simple GL-like theory with a 
finite number of spatial derivatives of $\Delta$, does, 
unfortunately, not exist for the present problem. Such an 
expansion may be performed for $\theta=0$, in the purely 
paramagnetic limit, near the tricritical point $T_{tri}$, where 
the order parameter gradient is small because the 
characteristic length $q^{-1}$ of the FFLO state diverges 
at $T_{tri}$. However, for finite $H_\perp$ a small characteristic 
length for order parameter variations does not exist 
in the relevant range of temperatures, and the spatial 
variation of $\Delta$ must be taken into account exactly. 
One might still hope that a GL theory with a finite 
number of derivatives, although not accurate, will be 
useful to predict the \emph{qualitative} behavior of the 
superconducting states near $H_{c2}$ correctly; bearing in 
mind for example the results of standard GL for type II
superconductivity. However, for the mixed 
orbital-paramagnetic pair-breaking phenomena under 
discussion, there is not even a single point on 
the temperature scale where a GL theory with a 
finite number of derivatives is valid.
Such a theory is only valid near $T_c$ where no FFLO state 
exists, or near $T_{tri}$ in the ``vicinity'' of 
the paramagnetic limit, i.e. for extremely large $n$. The 
latter region is inaccessible both from a numerical and 
a experimental point of view. In this context, it should 
also be noted that the final equilibrium structures do 
not show any continuity with regard to $n$. 

Fortunately, the present problem does not require solving 
the full set of Gorkov's equations because the simpler 
set of quasiclassical equations may be used instead, as 
pointed out by Bulaevskii\cite{BULATILT}. The large parallel 
component $H_\|$ of the applied magnetic field, acting 
only on the spins of the electrons, is exactly taken 
into account by the Zeeman term. Thus, with regard to 
this component no question, as to the validity of the 
quasiclassical approximation, arises. The magnitude of 
the perpendicular component $H_\perp$, on the other hand, 
must obey the usual quasiclassical condition $\hbar\omega_c<k_B T$, 
where $\omega_c=eH_\perp/mc$, or $\sin\theta (e\hbar H/mc) < k_BT$. Inserting 
the highest possible field $H=H_P$ in the latter relation,
one finds that the quasiclassical approximation holds 
indeed for not too low temperatures, $T/T_c>k_BT_C/E_F$, in 
the interesting range of tilt-angles $\theta<\theta_1$, where the 
new paramagnetic vortex states appear.

In most papers on paramagnetic pair-breaking and 
the FFLO state the influence of orbital pair-breaking is
completely neglected. This means, that the 
GL-parameter $\kappa$ tends to infinity and that all spatial 
variations of the magnetic field can be neglected. 
For three-dimensional superconductors this approximation 
implies that the orbital critical field is much higher
than the paramagnetic Pauli-limiting field. This is 
impossible to achieve\cite{MANKL2} for BCS-like 
superconductors, because the superconducing coherence 
length cannot be smaller than an atomic distance. It 
seems unlikely even for unconventional 
materials\cite{GMSBGSBSK} where many-body 
effects may lead to a strong renormalization of the 
input parameters. For the present 2D situation, 
the suppression of the orbital pair-breaking effect is 
entirely due to geometrical reasons, and no restriction on 
the value of $\kappa$ is required in order to reach the purely 
paramagnetic limit at parallel fields. Thus, keeping
all terms in the quasiclassical free energy related to 
spatial variations of the magnetic field, will allow us
to study type II superconductors with arbitrary $\kappa$ or 
even type I material. Large-$\kappa$ superconductors
show, however, still a practical advantage because of 
their larger critical angle $\theta_1$ [see Eq.~(\ref{eq:docan1})]. 

This paper is organized as follows. 
In section~\ref{sec:quclze} 
Eilenbergers quasiclassical equations generalized with 
regard to a Zeeman coupling term, as well as the 
corresponding 
free energy functional, are reported. The expansion of the
free energy near the upper critical field, for a general 
2D quasi-periodic state, is treated in 
section~\ref{sec:entucf}. Two limiting cases of the 
analytical results, the GL limit and the structure of 
the ordinary vortex lattice, are reported in appendices. 
The numerical results for the paramagnetic vortex states, 
at finite perpendicular field, are reported and discussed 
in section~\ref{sec:rffpf}. The structure of the FFLO 
state, for the special case of vanishing perpendicular 
field, is reconsidered in the present quasiclassical 
framework in section~\ref{sec:sotfs}. The 
non-trivial transition $\theta\to0$ (or $n\to\infty$) to the purely 
paramagnetic limit is analyzed in section~\ref{sec:tttppr}. 
An explanation for the increase in $n$, in terms of 
the finite momentum of the Cooper pairs in the 
paramagnetic vortex states, is also reported in this 
section. The results are summarized in the final 
section~\ref{sec:concln}.

\section{\label{sec:quclze} Quasiclassical equations 
with Zeeman term}  

We need a weak-coupling, clean-limit version of the
quasiclassical theory\cite{EILE1,LAROVC}, which contains 
all terms related to the coupling of the electron's spins 
to an external magnetic field. A general quasiclassical 
theory which covers Zeeman coupling 
has been published by Alexander et.al\cite{AORT}.
The $4\times4$ Green` s function matrix appearing in this work
may be considerably simplified for the present situation. 
Since we neglect spin-orbit coupling, the direction of the 
magnetic induction $\vec{B}$ in spin-space may be chosen 
independently from the direction of $\vec{B}$ in 
ordinary space; we adopt the usual choice of $\vec{B}$ 
being parallel to the $z-$direction in spin space. Then, 
only six essential Green's functions remain, which are 
denoted by
\begin{eqnarray*}
  \label{eq:twogroupsgf}
f_{(+)}  &=& f+f_3\hspace{1cm}f_{(-)} =f-f_3     \nonumber\\ 
f_{(-)}^{+} &=& f^+-f_3^+\hspace{0.6cm}f_{(+)}^{+} = f^++f_3^+ \nonumber\\ 
g_{(+)}  &=& g+g_3 \hspace{1cm}g_{(-)} = g-g_3
\mbox{.}    
\end{eqnarray*}
Here, $f,\,f^+,\,g$ denote the Greens functions in the 
absence of Zeeman coupling, and $f_3,\,f_3^+,\,g_3$ are the 
additional Green`s function components in the the 
$z-$direction of spin space. The three equations for the 
left group $f_{(-)},\,f_{(+)}^{+},\,g_{(-)}$ are decoupled from the 
three equations for the right group $f_{(+)},\,f_{(-)}^{+},\,g_{(+)}$ and 
differ only by a negative sign in front of the magnetic 
moment $\mu \doteq  \hbar |e|/(2mc)$ of the electron. Also, for each 
group a separate normalization condition $g_{(+)}^2+f_{(+)}f_{(-)}^+=1$ 
and $g_{(-)}^2+f_{(-)}f_{(+)}^+=1$ respectively, exists. Therefore, it 
is convenient to introduce Greensfunctions $f,\,f^+,\,g$,
defined by 
\begin{eqnarray*}
  \label{eq:convgreenf}
f(\vec{r},\vec{k},\omega_s)  &=& f_{(-)}(\vec{r},\vec{k},\omega_) \mbox{,} 
\nonumber\\ 
f^{+}(\vec{r},\vec{k},\omega_s) &=& f_{(+)}^{+}(\vec{r},\vec{k},\omega) \mbox{,} 
\nonumber\\ 
g(\vec{r},\vec{k},\omega_s)  &=& g_{(-)}(\vec{r},\vec{k},\omega)
\mbox{,}    
\end{eqnarray*}
which are functions of the  spatial variable
$\vec{r}$, the quasiparticle wave-number $\vec{k}$, 
and the complex variable $\omega_s=\omega+\imath\mu B$. The 2D 
variable $\vec{r}$ denotes positions in
the conducting ($x,y$)-plane. The real variable $\omega$ takes 
the values of the Matsubara frequencies $\omega_l=(2l+1)\pi k_BT$;
the Matsubara index $l$ will not always be written down 
explicitly. The second group of Green`s functions 
$f_{(+)},\,f_{(-)}^{+},\,g_{(+)}$ may be expressed by similar 
relations in terms of $f,\,f^+,\,g$ if $\omega_s$ is replaced by
$\omega_s^{\ast}$.  

Using the Green`s functions $f,\,f^+,\,g$, the quasiclassical 
equations  with Zeeman coupling become formally similar to the
quasiclassical equations without spin terms. The nonlinear 
transport equations for $f,\,f^+$ are given by  

\begin{equation}
  \label{eq:quasiffp}
\begin{split}
\left[ 2\omega_s+\hbar \vec{v}_F(\vec{k})\vec{\vartheta}_r \right] 
f(\vec{r},\vec{k},\omega_s) & =  2\Delta(\vec{r}) 
g(\vec{r},\vec{k},\omega_s)\mbox{,} \\
\left[ 2\omega_s - \hbar \vec{v}_F(\vec{k}) \vec{\vartheta}_r^{\ast}\right] 
f^{\it+}(\vec{r},\vec{k},\omega_s) & =  2\Delta^\ast(\vec{r}) 
g(\vec{r},\vec{k},\omega_s)
\mbox{,}
\end{split}     
\end{equation}

where the Green`s function g is given by the normalization 
condition
\begin{equation}
  \label{eq:quasig}
g(\vec{r},\vec{k},\omega_s) = \left( 1-
f(\vec{r},\vec{k},\omega_s) 
f^{\it+}(\vec{r},\vec{k},\omega_s) \right)^{1/2} \mbox{.}
\end{equation}

Here, $\vec{v}_F(\vec{k})$ denotes the Fermi velocity and 
$\vec{\vartheta}_r$ is the gauge-invariant derivative defined by 
$\vec{\vartheta}_r=\vec{\nabla}_r-\imath(2e/\hbar c) \vec{A}$. The order parameter 
$\Delta$ and the vector potential $\vec{A}$ must be determined 
selfconsistently. 

The self-consistency equation for $\Delta$ is given by
\begin{widetext}
\begin{equation}
  \label{eq:scop}
\left(2 \pi k_B T  \sum_{l=0}^{N_D} \frac{1}{\omega_l}+ 
\ln\left(T/T_c\right) \right) 
\Delta(\vec{r}) = \pi k_B T  \sum_{l=0}^{N_D} 
\oint d^2k' \, \left[ f(\vec{r},\vec{k}^\prime,\omega_s) +
f(\vec{r},\vec{k}^\prime,\omega_s^\ast ) \right] 
\mbox{,}
\end{equation}
\end{widetext} 
where $N_D$ is the cutoff index for the Matsubara sums.
The self-consistency equation for $\vec{A}$ is Maxwell`s 
equation
\begin{equation}
  \label{eq:scvecpot}
\begin{split}
&\vec{\nabla}_r \times \left( 
\vec{B}(\vec{r}) + 4\pi \vec{M}(\vec{r}) \right)= \\
&\frac{16\pi^2ek_BTN_F}{c} \sum_{l=0}^{N_D} 
\oint \frac{d^2k'}{4\pi} \, \vec{v}_F(\vec{k}^\prime)
\Im g(\vec{r},\vec{k}^\prime,\omega_s ) \mbox{,} 
\end{split}
\end{equation}
where $N_F$ is the normal-state density of states at the
Fermi level. The r.h.s. of Eq.~(\ref{eq:scvecpot}) is the 
familiar (orbital) London screening current while the 
magnetization $\vec{M}$ is a consequence of the magnetic 
moments of the electrons and is given by   
\begin{equation}
  \label{eq:magn}
\begin{split}
\vec{M}(\vec{r})=
&2\mu^2N_F\vec{B}(\vec{r}) \\
&-4\pi k_BTN_F \mu \sum_{l=0}^{N_D} \oint  \frac{d^2k'}{4\pi} 
\Im g \frac{\vec{B}}{B} 
\mbox{,} 
\end{split}
\end{equation}  
The first term on the r.h.s. of Eq.~(\ref{eq:magn}) 
is the normal state spin polarization. The second term
is is a spin polarization due to quasiparticles in the 
superconducting state.  

The following symmetry relations hold for solutions
of Eqs.~(\ref{eq:quasiffp},\ref{eq:quasig},\ref{eq:scop},\ref{eq:scvecpot})
\begin{equation}
  \label{eq:gensymrel}
\begin{split}
&g^\star(\vec{r},-\vec{k},\omega_s^\ast)=g(\vec{r},\vec{k},\omega_s) \mbox{,} \\
&f^+(\vec{r},\vec{k},\omega_s)=f^\star(\vec{r},-\vec{k},\omega_s^\ast) \mbox{,} \\
&g(\vec{r},-\vec{k},-\omega_s)=-g(\vec{r},\vec{k},\omega_s) \mbox{,} \\
&f(\vec{r},-\vec{k},-\omega_s)=f(\vec{r},\vec{k},\omega_s) \mbox{,} \\
&f^+(\vec{r},-\vec{k},-\omega_s)=f^+(\vec{r},\vec{k},\omega_s) \mbox{,} \\
\end{split}  
\end{equation}
which have been extensively used in the calculations 
described in the next sections.

The quasiclassical 
equations~(\ref{eq:quasiffp},\ref{eq:scop},\ref{eq:scvecpot})
may be derived as Euler-Lagrange equations of the Gibbs 
free energy functional $G$, which is given by 
\begin{widetext}
\begin{equation}
\label{eq:gibbsfe}
\begin{split}  
G=&\frac{1}{F_p}\int \,d^3r\Bigg[ 
\frac{\vec{B}^2}{8\pi}-\mu^2N_F\vec{B}^2 - \frac{\vec{B}\vec{H}}{4\pi}+
\\&N_F\left(\pi k_B T  \sum_{l=-\infty}^{+\infty} \frac{1}{|\omega_l|}+ 
\ln\left(T/T_c\right) \right)|\Delta|^2 
-\pi k_BTN_F \sum_{l=-\infty}^{+\infty} \oint  \frac{d^2k}{4\pi}\,I(\vec{r},\vec{k},\omega_s)
\Bigg]
\mbox{.}    
\end{split}  
\end{equation}
\end{widetext}
The area of the sample is denoted by $F_p$ and the 
$k-$dependent quantity $I$ is given by 
\begin{equation*}
  \label{eq:kdinteg}
\begin{split}
&I(\vec{r},\vec{k},\omega_s)=\Delta f^+ +\Delta^{\star } f + 
\left(g-\frac{\omega_l}{|\omega_l|}\right)\cdot \\
&\left[
\frac{1}{f}\left(\omega_s+\frac{\hbar\vec{v}_F}{2}\vec{\partial}_r\right)f+
\frac{1}{f^{+}}\left(\omega_s-\frac{\hbar\vec{v}_F}{2}\vec{\partial}_r^{\star }\right)f^+
\right]
\mbox{.}
\end{split}     
\end{equation*}

An important reference state for the present problem 
is the purely paramagnetically limited homogeneous 
superconducting state, which is realized for our 
2D superconductor if the magnetic field 
is exactly parallel to the conducting planes. In this 
case, the vector potential and the gradient terms in the 
transport equations may be omitted. At $T=0$ the free 
energy difference between the superconducting and 
normal-conducting states may be derived analytically. 
It is given by 
\begin{equation}
  \label{eq:homgibbsdiff}
G_s-G_n=N_F \left(\mu^2H^2-\Delta_0^2/2\right)
\mbox{,}
\end{equation}
and vanishes at the Pauli critical field $H_P$. For higher
$T$ the self-consistency equation for the gap must be solved 
numerically, yielding agreement with previous results.
\cite{SARMA,BURKRAIN}. Let us investigate the magnetic 
response in this purely paramagnetic limit. It is 
neglected in most theoretical treatments, but is of 
particular interest if the influence of finite values 
of the GL-parameter $\kappa$ is to be taken into account. To 
obtain the magnetization due to the spins,
the coupled self-consistency 
equations~(\ref{eq:scop},\ref{eq:scvecpot}) have to  
be solved. Using dimensionless quantities defined in 
appendix~\ref{sec:sou} the  gap equation takes the form 
\begin{equation}
  \label{eq:gapdiml}
\ln t-t\sum_{l=0}^{N_D}\left[ \left(\frac{1}{\sqrt{|\Delta|^2+
\left(\omega_l+\imath\mu B \right)^2}}+\mbox{c.c.} \right)-
\frac{2}{\omega_l}  \right]=0
\mbox{,}
\end{equation}
while Maxwell's equation reduces to 
\begin{equation} 
  \label{eq:magpdiml} 
B-H=\frac{\mu}{\tilde{\kappa}^2}\left(\mu B-2t\sum_{l=0}^{N_D}
\Im \frac{\omega_l+\imath\mu B}{\sqrt{|\Delta|^2 + \left(\omega_l+\imath\mu B \right)^2}}
\right)
\mbox{.}
\end{equation}
Note that the orbital screening current 
[the r.h.s. of Eq.~(\ref{eq:scvecpot})] is completely 
absent for the plane-parallel field direction. At 
$T=0$ the r.h.s. of Eq.~(\ref{eq:magpdiml}) vanishes 
exactly. This means that the normal state spin
polarization [first term on the r.h.s. of
Eq.~(\ref{eq:magpdiml}] is exactly canceled by the spin
polarization due to the superconducting quasiparticles
[second term on the r.h.s. of Eq.~(\ref{eq:magpdiml})].
The numerical solution shows that the quasiparticle 
polarization decreases with increasing $T$ and vanishes 
at $\Delta=0$, where the magnetic behavior of the 
normal-conducting state is recovered. 

In the rest of this paper dimensionless quantities 
as introduced by Eilenberger will be used. These 
quantities are listed in appendix~\ref{sec:sou}. 
Any exceptions will be mentioned explicitly.

\begin{figure}[htbp] 
\begin{center}\leavevmode
\includegraphics[width=8cm]{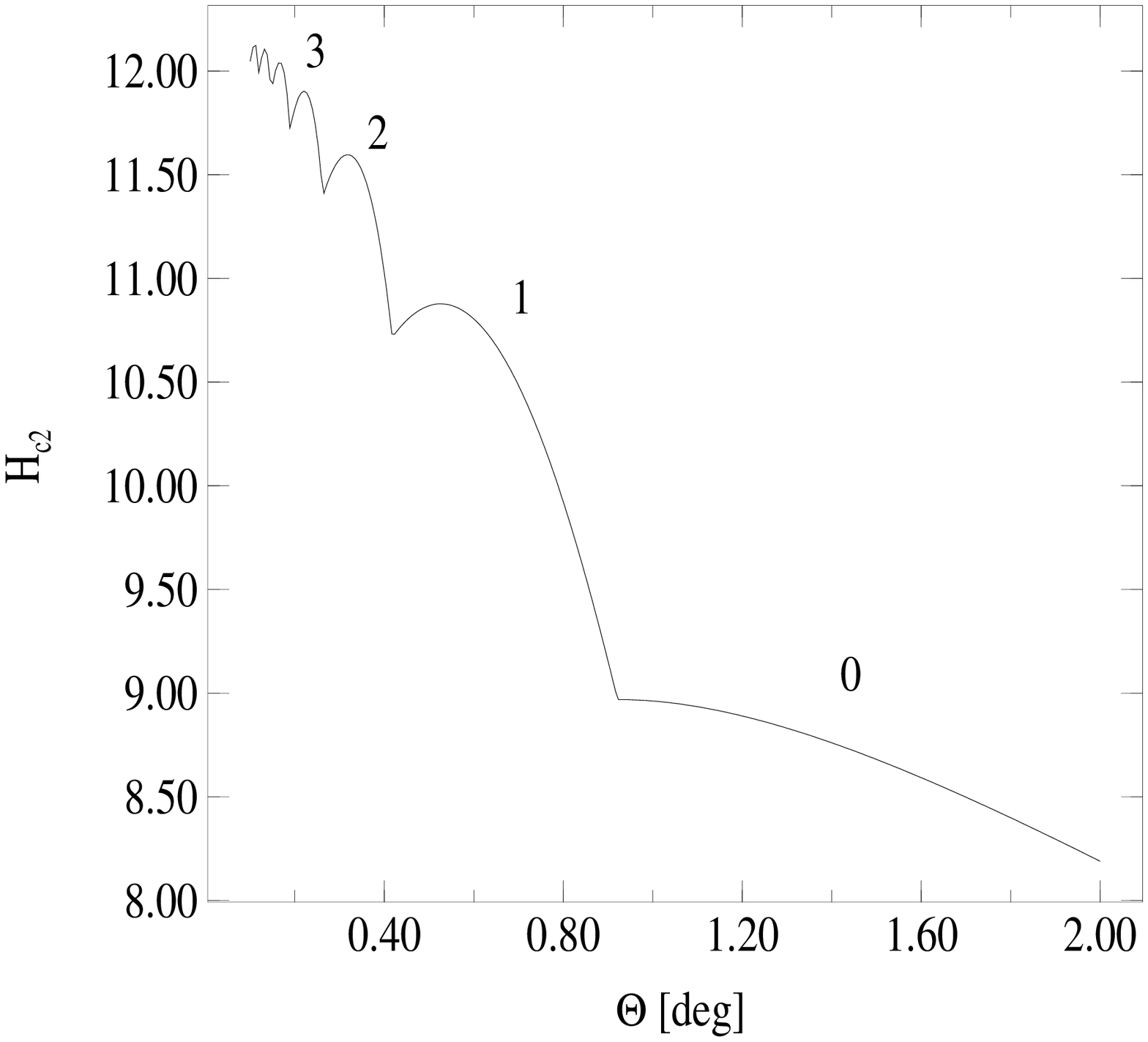}
\caption{\label{fig:phabou} Phase boundary of the 
superconducting state at $t=0.1$ for tilt angles $\Theta$ 
between $0.1$ and $2.0$ using a value $\mu=0.04$  for the 
dimensionless magnetic moment of the electron. The numbers 
$0,1,2,\ldots$ are Landau quantum numbers characterizing the 
individual pieces of the curve.}
\end{center} 
\end{figure}

In the next sections the stable order parameter structure 
of a 2D superconductor in the vicinity of the phase 
boundary will be investigated. The phase boundary $H_{c2}(T)$ 
itself is given by the highest solution of the 
equation\cite{SHIMARAIN}
\begin{equation}
  \label{eq:hc2equ}
\begin{split}
0=\ln t + & t \int_0^{\infty}\textrm{d}s\,\frac{1- \textrm{e}^{-\omega_Ds}} {\sinh st}\Big[1 \\
-& \cos(\mu H s)\textrm{e}^{-H_{\perp}s^2/4}L_n(H_{\perp}s^2/2)
 \Big]
\mbox{,}
\end{split}
\end{equation}
where the integer $n=0,1,2,\ldots$ is Landau's quantum number, 
$\omega_D$ is the Debye frequency, and $L_n$ is a Laguerre 
polynomial\cite{GRADRYSH} of order $n$. A typical phase 
boundary is shown in Fig.~\ref{fig:phabou}. Each piece of 
the nonmonotonic $H_{c2}$ curve is characterized by a 
single value of $n$.           
An infinite number of eigenstates $\phi_{n,k}$ exists, belonging 
all to the same, highly degenerate eigenvalue $n$. For 
the present gauge, these are given by
\begin{equation}
  \label{eq:bequphe}
\begin{split}
\phi_{n,k}(\vec{r})=&A\frac{(-1)^n}{\sqrt{n!}}\textrm{e}^{\imath kx}
\textrm{e}^{-\frac{H_{\perp}}{2}\left(y-\frac{k}{H_{\perp}}\right)^2}
\cdot \\
&He_n\left(\sqrt{2H_{\perp}}\left[y-\frac{k}{H_{\perp}} \right] \right)
\mbox{,}
\end{split}
\end{equation} 
where $k$ is a real number and $He_n$ is a Hermite 
polynomial\cite{GRADRYSH} of order $n$. The 
functions~(\ref{eq:bequphe}) are orthogonal and normalized,       \begin{equation}
  \label{eq:orthrel}
(\phi_{n,k},\phi_{m,l})=\delta_{n,m}\delta(k-l)
\mbox{,}
\end{equation}
if the amplitude $A$ in Equ.~(\ref{eq:bequphe}) is chosen 
according to
\begin{equation}
  \label{eq:normefho}
A=\frac{1}{R_0}\left(\frac{H_{\perp}}{\pi L_x^2} \right)^{1/4}
\mbox{,}
\end{equation}
where $L_x$ is the size of the system in $x-$direction and 
$R_0$ is defined in appendix~\ref{sec:sou}. The 
gap, for the portion of the $H_{c2}$-curve characterized 
by $n$, is a linear combination of all $\phi_{n,k}$ belonging 
to this $n$. The harmonic oscillator 
eigenfunctions~(\ref{eq:bequphe}) are extensively 
used in the theory of the quantum Hall\cite{PRANSTE} 
effect and many other topics in the quantum theory of 
a charged particle in a magnetic field. 
\section{\label{sec:entucf} Free energy expansion near 
the upper critical field}
We assume that the transition between the superconducting
and normal-conducting states at the upper critical field 
$H_{c2}$ will be of second order for arbitrary tilt-angle $\theta$. 
Then, the order parameter $\Delta$, or more precisely its 
amplitude $\epsilon$ may be used as a small parameter 
for expanding the free energy $G$ in the vicinity 
of $H_{c2}$. We keep terms up to fourth order in $\epsilon$  
and all orders in order parameter derivatives and 
determine the energetically most favorable order 
parameter structure near $H_{c2}$. Similar calculations
for the ordinary vortex lattice, corresponding to the 
case of large $\Theta$ of the present arrangement, have 
been performed by Eilenberger\cite{EILE2} and by 
Rammer and Pesch\cite{RAMPESCH}. No special assumptions 
on the order parameter structure, such as the number
of zeros per unit cell, will be made. We only assume 
that the order parameter is quasi-periodic on a 
2D lattice, with an arbitrary unit cell, 
characterized by the length of the two basis vectors 
and the angle between them. The free energy will 
be minimized with respect to these unit cell 
parameters.

\subsection{Order parameter}
\label{subsec: orderpar}

Let the unit vector $\vec{a}$ of our elementary cell be
parallel to the $x-$axis, $\vec{a}=a\vec{e}_x$. The angle 
between $\vec{a}$ and the second unit vector $\vec{b}$
is denoted by $\alpha$. To construct a quasi-periodic order
parameter near $H_{c2}$, exactly the same method as used by 
Abrikosov\cite{ABRI1}, for the case $n=0$, may be applied. 
The result is given by the following linear combination 
of a subset of the basis functions~(\ref{eq:bequphe})
\begin{eqnarray}
 \label{eq:opdimlos}
\Delta_n(\vec{r})&=&AC_n\sum_{m=-\infty}^{m=+\infty}\exp\left(-\imath \pi \frac{b}{a}m(m+1) \cos\alpha  \right) \times\nonumber \\ 
& &\exp\left(\imath \frac{2\pi}{a}m x  \right) h_n\left(y-mb\sin\alpha \right)
\mbox{,}
\end{eqnarray} 
where
\begin{equation*}
h_n(z)=\frac{(-1)^n}{\sqrt{n!}}
\textrm{e}^{-\frac{\bar{B}_{\perp}}{2}z^2}
He_n\left(\sqrt{2\bar{B}_{\perp}}z \right)
\mbox{.}
\end{equation*}
This order parameter\cite{AKDOGINO,RIESCHSCH,KLRAISHI} is 
not invariant under translations 
$\vec{r}\to\vec{r}^{\,'}=\vec{r}+n\vec{a}+m\vec{b}$ but acquires 
phase factors for each elementary translation, which are 
uniquely defined within a fixed gauge. 
Surrounding a unit cell in anti-clockwise direction, these 
phase changes add up to a total factor of $\exp \imath2\pi$, 
i.e. each unit cell carries a single flux quantum $\Phi_0$ . 
We shall use this assumption of a single flux quantum per 
unit cell, which is written as $\bar{B}_{\perp}ab\sin\alpha=
2\pi$ in the present units, throughout this paper. Preliminary 
calculations\cite{KLEINUNP} show that states with two flux 
quanta per unit cell have higher free energy and can be 
excluded. Also, a preference for multi-quanta vortices seems 
unlikely in the present situation, where the single flux 
quantum state is stable at large $\Theta$, while the total 
flux decreases to zero as $\Theta\to0$.

The order parameter~(\ref{eq:opdimlos}) describes a flux 
line lattice where the Cooper pair states belong to 
arbitrary Landau quantum numbers $n$, depending on the tilt 
angle $\Theta$. As is well known, the pairing states for the 
ordinary vortex state belong to the lowest Landau level $n=0$. 
The present shift to higher Landau levels is, of course, 
related to the large paramagnetic pair-breaking field $H_{\parallel}$ 
as will be discussed in more detail in section~\ref{sec:tttppr}.

The coefficient $C_n$ in Eq.~(\ref{eq:opdimlos}) may be expressed
by the spatial average of the square of the order parameter, 
using the relation
\begin{equation}
  \label{eq:amplclim}
\langle|\Delta_n|^2\rangle=\frac{1}{F_p}|C_n|^2\sum_{m=-M/2}^{+M/2}1
\mbox{,}
\end{equation}
where $F_p$ is the area of the sample. The spatial average 
over the unit cell area $F_c=ab\sin\alpha$ is defined in 
appendix ~\ref{sec:sou}. For later use, when 
performing the limit $\Theta\to0$ in section~\ref{sec:tttppr}, we 
assumed in Eq.~(\ref{eq:amplclim}) that the area of the 
superconducting plane is finite and that the number of unit 
cells in one direction is $M$. At the end of the following
calculation, $C_n$ will be fixed according to the requirement 
$\langle|\Delta_n|^2\rangle=1$ and an infinitesimal amplitude $\epsilon$ 
will be attached in front of each power of $\Delta_n$. 

A useful quantity is the square of the order parameter 
modulus, which may be written in the form 
\begin{equation}
  \label{eq:opmodsqare}
|\psi_{n}|^2(\vec{r})= \sum_{l,j} (\psi_{n}^2)_{l,j}\textrm{e}^{\imath \vec{Q}_{l,j}\vec{r}}
\mbox{.}
\end{equation}
The Fourier coefficients $(\psi_{n}^2)_{l,j}$ are given by 
\begin{equation}
  \label{eq:opmsfc}
(\psi_{n}^2)_{l,j}=(-1)^{lj} \textrm{e}^{-\imath \pi l \frac{b}{a}\cos\alpha } 
\mathrm{e}^{-x_{l,j}/2} L_n(x_{l,j})
\mbox{,}
\end{equation}
where $x_{l,j}$ is defined by Eq.~(\ref{eq:affs6k}). The 
order parameter $\psi_n$ is proportional to $\Delta_n$ but with an
amplitude chosen according to $\langle |\psi_{n}|^2 \rangle =1$.  
It is instructive to compare Eq.~(\ref{eq:opmodsqare}) 
with the local magnetic field reported later in 
subsection~\ref{subsec:locinduct}.   

\subsection{General aspects of the expansion}
\label{subsec:genasexp}
A fourth order expansion of $G$ requires first and third order
contributions in the Greens functions $f,\, f^+$. We use the 
notation   
\begin{equation} 
  \label{eq:gfexp}
f=f^{(1)}+f^{(3)},\hspace{0.5cm}f^{+}=f^{+(1)}+f^{+(3)}
\mbox{,}
\end{equation}
where $f^{(1)}$ and $f^{(3)}$ are the contributions of order 
$\epsilon^{1}$ and $\epsilon^{3}$ respectively. A consistent treatment of 
the magnetic field terms\cite{EILE3,KLEIN1} requires a 
separation of $\vec{B}$ and $\vec{A}$  according to 
\begin{equation} 
  \label{eq:magfspl}
\vec{B}(\vec{r})=\bar{\vec{B}}+\vec{B}_1(\vec{r}),\hspace{0.3cm}
\vec{A}(\vec{r})=\bar{\vec{A}}(\vec{r})+\vec{A}_1(\vec{r})
\mbox{,}
\end{equation}
where $\bar{\vec{B}}$ is the spatially constant magnetic 
induction, $\vec{B}_1(\vec{r})$ is the $\vec{r}-$dependent  
deviation from $\bar{\vec{B}}$ and $\bar{\vec{A}}(\vec{r})$,
$\vec{A}_1(\vec{r})$ are the corresponding vector potentials.
An evaluation of the magnetic field terms in $G$ requires the 
leading order in $\vec{B}_1(\vec{r})$, which is $\epsilon^2$: 
$\vec{B}_1\approx\vec{B}_1^{(2)}$. The spatially constant quantity
$H_{c2}-\bar{B}$, where $\bar{B}=|\bar{\vec{B}}|$, is small of 
order $\epsilon^2$. The whole expansion in $\epsilon$ will be done keeping 
$\bar{B}$ fixed; at the end of the calculation, the 
Gibbs free energy $G$ will be minimized with respect 
to the order parameter amplitude $\epsilon$ and the induction 
$\bar{B}$. The calculation can be seen as an extension 
of Abrikosov's classical work\cite{ABRI1} to arbitrary 
temperatures below $T_c$.

Let us choose the coordinate system in such a way that the 
magnetic field lies in the $(y,z)-$plane. Then, the 
induction $\vec{B}(\vec{r})$ (and the external field  
$\vec{H}$) may be split according to 
\begin{equation}
  \label{eq:bgeomspl}
\vec{B}(\vec{r})=B_{\parallel}(\vec{r})\vec{e}_y+B_{\perp}(\vec{r})\vec{e}_z
\mbox{,}
\end{equation}
in perpendicular and parallel components $B_{\perp},\,B_{\parallel}$. The 
corresponding vector potentials are denoted by 
$\vec{A}_{\perp},\,\vec{A}_{\parallel}$. In order to fix the gauge we may 
employ here essentially the same method as used before in 
numerical calculations on the vortex lattice without Zeeman 
coupling\cite{KLEIN1,KLEIN2}. The gauge conditions which 
fix $\vec{A}_1$ are given by\cite{EILE3} 
\begin{equation}
  \label{eq:gcfa1}
\frac{\partial\vec{A}_1}{\partial\vec{r}}=0,\;\int d^2r\,\vec{A}_1=0,\; 
\vec{A}_1\,\mbox{periodic}
\mbox{.}
\end{equation}
The vector potential $\bar{\vec{A}}$ describing the 
average value $\bar{\vec{B}}$  of the induction is chosen 
according to 
\begin{equation}
  \label{eq:vpsci}
\bar{\vec{A}}(\vec{r})=\left(\bar{B}_{\parallel}z-
\bar{B}_{\perp}y \right)\vec{e}_x
\mbox{.}
\end{equation}
The first term in Eq.~(\ref{eq:vpsci}) can be omitted 
in the gauge invariant derivatives of 
Eq.~(\ref{eq:quasiffp}) since no $z-$dependence 
exists in our 2D system. Thus, the orbital 
pair-breaking contribution in the transport equations 
consists of the sum of the second term $\propto\bar{B}_{\perp}$ in 
Eq.~(\ref{eq:vpsci}) and the $\vec{r}-$dependent part 
$\vec{A}_1$ (only the perpendicular component 
of $\vec{A}_1$ is relevant here). The (large) parallel 
component $\bar{B}_{\parallel}$, on the other hand, enters the spin 
pair-breaking term, which is proportional to 
$B(\vec{r})=(B_{\parallel}^2(\vec{r})+B_{\perp}^2(\vec{r}))^{1/2}$. 
Eqs.~(\ref{eq:gcfa1},\ref{eq:vpsci}) fix the gauge, i.e. 
allow a unique determination of $\vec{A}$ in terms of 
$\vec{B}$. While $|\Delta|^2$ and $\vec{B}$ are periodic, i.e. 
invariant under translations between equivalent points in 
the 2D structure, $\Delta$ and $\vec{A}$  are 
only quasiperiodic, i.e. they differ by phase factors and 
a change in gauge respectively. The phase factors are 
fixed within a given gauge and may be calculated using 
Eq.~(\ref{eq:vpsci}).

As a first step in the expansion of $G$, the 
Green's function $g$ is eliminated in favor of $f,\,f^+$
by means of the relation   
\begin{equation*}
  \label{eq:gexpinf}
g=1-\frac{f f^+}{2}-\frac{f^2(f^+)^2}{8}+\ldots
\mbox{,} 
\end{equation*} 
which is valid for small $\Delta$. Second, the gradient terms
in $G$ may be eliminated with the help of the 
transport equations~(\ref{eq:quasiffp}). Then, the 
(dimensionless) Gibbs free energy takes the form   
\begin{widetext}
\begin{equation}
\label{eq:gisfedl}
\begin{split}
G=\frac{1}{F_p}\int \,d^3r\Bigg[ 
\tilde{\kappa}^2\left(\vec{B}-\vec{H} \right)^2-\mu^2\vec{B}^2 +
& \left(\ln t+2\sum_{l=0}^{\infty}\frac{1}{2l+1}\right)|\Delta|^2 \\ -
& \frac{t}{2} \sum_{l=0}^{\infty} \left[
\Delta \overline{f^+}+\Delta^{\star }\overline{f}+
\frac{1}{4}\left(
\Delta \overline{f(f^+)^2} + \Delta^{\star } \overline{f^2f^+} \right)+\mbox{c.c.}
 \right]
\Bigg]
\mbox{,}
\end{split}    
\end{equation}
\end{widetext} 
where the bar denotes a Fermi surface average as defined 
in appendix~\ref{sec:sou}.

Inserting the expansions~(\ref{eq:gfexp}),(\ref{eq:magfspl})
in the free energy~(\ref{eq:gisfedl}) and collecting terms
of the same order in $\epsilon$, $G$ takes the form 
\begin{equation}
  \label{eq:gsplititt}
G=\bar{G}\,+G^{(2)}+G^{(4)}
\mbox{,}
\end{equation}
where the terms $\bar{G},\,G^{(2)}$, and $G^{(4)}$ denote the 
free energy contributions of order $\epsilon^0,\,\epsilon^2$ and $\epsilon^4$ 
respectively. The term $\bar{G}$ is given by
\begin{equation}
  \label{eq:zerogfe}
\bar{G}=\tilde{\kappa}^2 \left(\bar{\vec{B}}-\vec{H} \right)^2-
\mu^2\bar{\vec{B}}^2
\mbox{.}
\end{equation}
We will first simplify the quantities $G^{(2)}$ and $G^{(4)}$ and 
then calculate the minimum of $G$ with respect to the 
amplitude $\epsilon$ and the induction $\bar{B}$.

\subsection{Second order contribution}
\label{subsec:secordcon}
The second order contribution to the Gibbs free energy 
is given by   
\begin{equation}
\begin{split}
  \label{eq:s2octg}
G^{(2)}&=\frac{1}{A}\int \,d^3r\Bigg[  
\left(\ln t+2\sum_{l=0}^{\infty}\frac{1}{2l+1}\right)|\Delta|^2\\
&- \frac{t}{2} \sum_{l=0}^{\infty} \left(
\Delta \overline{f^{+(1)}}+\Delta^{\star }\overline{f^{(1)}}+\mbox{c.c.}
\right)\Bigg]
\mbox{.}
\end{split}
\end{equation}
To calculate the lowest order Green's function 
only contributions of order $\epsilon^0$, namely the 
spatially constant part $\bar{B}=\left(\bar{B}_{\parallel }^2+
\bar{B}_{\perp}^2 \right)^{1/2}$ of the induction and the lowest order 
vector potential $\bar{\vec{A}}$, have to be taken into 
account in Eq.~(\ref{eq:quasiffp}). The resulting equation
for $f^{(1)}$ is given by  
\begin{equation}
  \label{eq:fowff}
[\omega_l+\imath \mu \bar{B}+\hat{k}\vec{\partial}_{r}^{(0)} ]f^1=\Delta
\mbox{,}
\end{equation} 
where $\vec{\partial}_{r}^{(0)}=\frac{\partial}{\partial\vec{r}}+\imath\bar{B}_{\perp}y\vec{e}_x$.
To proceed, we use well-known methods\cite{HELFWERT} and 
solve first the eigenvalue problem of the operator 
$\hat{k}\vec{\partial}_{r}^{(0)}$. The solution is given by
\begin{equation}
  \label{eq:dfaevoaop}
\hat{k}\vec{\partial}_{r}^{(0)} f_{\hat{k},\vec{p}}(\vec{r})=E_{\hat{k},\vec{r}} f_{\hat{k},\vec{p}}(\vec{r})
\mbox{,}
\end{equation}
with the eigenvalues $E_{\hat{k},\vec{r}} =\imath\hat{k}\vec{p}$ and the 
eigenfunctions 
\begin{equation}
\label{eq:efofhwp}
\begin{split}
f_{\hat{k},\vec{p}}(\vec{r})&=\textrm{exp}  \bigg[ \imath
\frac{\bar{B}_{\perp}}{2}\left(x \hat{k}_x+y \hat{k}_y \right)
\left( x \hat{k}_y-y \hat{k}_x \right) \\
&-\imath\bar{B}_{\perp}\frac{xy}{2}+\imath\vec{p}\vec{r}\;
\bigg]
\mbox{.}
\end{split}
\end{equation}
Using the completeness of this continuous set of 
eigenfunctions the differential operator on the 
l.h.s. of Eq.~(\ref{eq:fowff}) may be inverted and 
$f^{(1)}$ be represented in the form
\begin{equation}
\label{eq:arffifo}
f^{(1)}=\int\frac{\textrm{d}^2p}{4\pi^2}\,\int \textrm{d}^2r_1\,
\frac{f_{\hat{k},\vec{p}}(\vec{r})f_{\hat{k},\vec{p}}^{\star}(\vec{r}_1)}{\omega_l+\imath \mu \bar{B}+
\imath \hat{k}\vec{p}}\;\Delta(\vec{r}_1) 
\mbox{.}
\end{equation}
Representing the denominator in Eq.~(\ref{eq:arffifo}) 
by  means of the identity 
\begin{equation}
\label{eq:asidtt}
\frac{1}{r}=\int_0^{\infty}\textrm{d}s\, \textrm{e}^{-sr}
\mbox{}
\end{equation}
as an additional integral, both the $\vec{p}-$integration 
and the $\vec{r}_1-$integration may be performed analytically 
and the solution of Eq.~(\ref{eq:fowff}) takes the form   
\begin{equation}
\label{eq:fsffotp} 
\begin{split}
&f^{(1)}(\hat{k},\omega_s,\vec{r}) =  
\int_0^{\infty}\textrm{d}u\,\textrm{e}^{-u\omega_s}\cdot  \\
&\textrm{exp}\left[
\imath \frac{\bar{B}}{2}
\left(-2uy\hat{k}_x+u^2\hat{k}_x\hat{k}_y \right)
\right]\,\Delta^{\star}(\vec{r}-u\hat{k})
\mbox{.}
\end{split}
\end{equation}
The first order solution for $f^+$ is given by 
$f^{+(1)}(\hat{k},\omega_s,\vec{r})=f^{(1)\star }(-\hat{k},\omega_s^{\star },\vec{r})$.

The evaluation of the remaining integrals may be greatly 
simplified by introducing the gap correlation function 
$V(\vec{r}_1,\vec{r}_2)$. In the present gauge it is defined by  
\begin{equation}
  \label{eq:defgapcorfun}
V(\vec{r}_1,\vec{r}_2)=\Delta(\vec{r}_1)\Delta^{\star}(\vec{r}_2)
\exp \left[\imath\frac{\bar{B}_{\perp}}{2}(x_1-x_2)(y_1+y_2) \right]
\mbox{.}
\end{equation}
Of particular importance are the Fourier coefficients 
$V_{l,j}(\vec{r})$, where $\vec{r}=\vec{r}_1-\vec{r}_2$. The 
precise definition and calculation of $V_{l,j}(\vec{r})$ 
is reported in appendix~(\ref{sec:gapcorfun}).

All terms in Eq.~(\ref{eq:s2octg}) containing first order 
Green`s functions may be expressed as integrals over a 
gap correlation function. The first of these takes the form 
\begin{equation}
  \label{eq:asrelwgc}
\Delta f^{+(1)}=\int_{0}^{\infty}\mathrm{d}u\,\mathrm{e}^{-u\omega_s}V^{\star}(\vec{r}+
u\hat{k},\vec{r})
\mbox{,}
\end{equation}
while the corresponding expression for $\Delta^{\star}f^{(1)}$ may be 
derived from~(\ref{eq:asrelwgc}) with the help of the 
symmetry relations~(\ref{eq:gensymrel}). To proceed,  
center of mass coordinates are introduced and a Fourier 
expansion of $V^{CM}(\vec{R},\vec{r})$ with regard to the 
variable $\vec{R}$ is performed, using the 
result~(\ref{eq:frffspa}) from appendix~\ref{sec:gapcorfun}. 
The remaining summations and integrations may be performed 
analytically\cite{GRADRYSH}. Collecting all terms one 
obtains the final result for the second order contribution  
\begin{equation}
  \label{eq:g2finres}
\begin{split}
G^{(2)}=&\langle|\Delta|^2\rangle\Big[
\ln t +  t \int_0^{\infty}\textrm{d}s\, \frac{1- \textrm{e}^{-\omega_Ds}} 
{\sinh st}\big[1 - \\
& \cos(\mu \bar{B} s)\textrm{e}^{-\bar{B}_{\perp}s^2/4}L_n(\bar{B}_{\perp}s^2/2)
 \big]
\Big]\mbox{.}
\end{split}
\end{equation}
While the order parameter expansion Eq.~(\ref{eq:opdimlos}),
which entered the calculation of $G^{(2)}$, depends on 
the lattice parameters $a,b,\alpha$, this dependence is  
absent in the final result, 
Eq.~(\ref{eq:g2finres}). The quantity $G^{(2)}$, characterizing 
the appearance of the superconducting instability, and 
not the detailed structure below it, does
only depend on the eigenvalue $n$. The relation $G^{(2)}=0$ 
agrees with the linearized gap equation~(\ref{eq:hc2equ}) 
used to calculate $H_{c2}$.

The technique used here to calculate $G^{(2)}$ 
will be generalized in the next subsection to 
evaluate the fourth order contribution to the free energy.   

\subsection{Fourth order contribution}
\label{subsec:fourthordcon}
The free energy contribution of order $\epsilon^4$ may be split, according to 
\begin{equation}
  \label{eq:splitfefo}
G^{(4)}=G^{(4)}_{N}+G^{(4)}_{M}
\mbox{,}
\end{equation}
in a nonmagnetic part $G^{(4)}_{N}$ and a magnetic part $G^{(4)}_{M}$.
In $G^{(4)}_{N}$ the spatially  constant induction $\bar{\vec{B}}$ and 
the corresponding vector potential $\bar{\vec{A}}(\vec{r})$
are used. The term $G^{(4)}_{M}$ collects all terms of order 
$\epsilon^4$ where deviations $\vec{B}_1(\vec{r})\approx \epsilon^2 $ (or the 
corresponding vector potential $\vec{A}_1(\vec{r})$) 
from the average induction $\bar{\vec{B}}$ are taken into 
account; for $\kappa\to\infty $, it becomes negigibly small.     

The nonmagnetic part $G^{(4)}_{N}$ is given by 
\begin{equation}
  \label{eq:nmpofgs}
G^{(4)}_{N}=G^{(4)}_{a}+G^{(4)}_{b}
\mbox{.}
\end{equation}
The term $G^{(4)}_{a}$ may be calculated using the solutions 
$f^{(1)},\,f^{+(1)}$ of order $\epsilon^1$, already obtained in 
subsection~\ref{subsec:secordcon},
\begin{equation}
\label{eq:g4nfirst}
G^{(4)}_{a}=-\langle
\frac{t}{8} \sum_{l=0}^{N_D} \left[
\Delta^{\star}\overline{f^{(1)2} f^{+(1)}}+ 
\Delta \overline{f^{(1)} f^{+(1)2}}+c.c.
  \right]\rangle
\mbox{.}
\end{equation}
The term $G^{(4)}_{b}$ requires the nonmagnetic parts 
$f^{(3)}_{N},\,f^{+(3)}_{N}$  of the third order Greens functions 
$f^{(3)},\,f^{+(3)}$,
\begin{equation}
\label{eq:g4nsecond}
G^{(4)}_{b}=-\langle
\frac{t}{2} \sum_{l=0}^{N_D} \left[
\Delta^{\star}\overline{f^{(3)}_{N}}+ 
 \Delta \overline{f^{+(3)}_{N}}+c.c.
\right]\rangle
\mbox{.}
\end{equation}    
The magnetic part $G^{(4)}_{M}$ is given by 
\begin{equation}
  \label{eq:mapofgs}
G^{(4)}_{M}= G^{(4)}_{c}+G^{(4)}_{d}
\mbox{.}
\end{equation}
The term $G^{(4)}_{c}$ is purely magnetic in origin, 
while the term $G^{(4)}_{d}$ contains the magnetic parts 
$f^{(3)}_{M},\,f^{+(3)}_{M}$  of the third order Greens functions,
\begin{eqnarray}
G^{(4)}_{c} &=& \langle(\tilde{\kappa}^2-\mu^2)\vec{B}_1^2\rangle  \label{eq:g4mpmp} 
\mbox{,} \\
G^{(4)}_{d} &=&-\langle
\frac{t}{2} \sum_{l=0}^{N_D} \left[
\Delta^{\star}\overline{f^{(3)}_{M}}+ 
 \Delta \overline{f^{+(3)}_{M}}+c.c.
\right]\rangle \label{eq:g4mfirt}
\mbox{.}
\end{eqnarray}  

In a next step, the terms $\vec{B}_1$ and $f^{(3)}=f^{(3)}_{N}+f^{(3)}_{M}$  
of order $\epsilon^2$ and $\epsilon^3$ respectively, must be calculated. The 
same method used in subsection~\ref{subsec:secordcon} to 
calculate $f^{(1)}$, by inverting the differential
operator on the l.h.s. of Eq.~(\ref{eq:fowff}), may  
be used here to obtain $f^{(3)}$. Using an operator notation for 
brevity, the sum of $f^{(3)}_{N}$ and  $f^{(3)}_{M}$ may be written as    
\begin{eqnarray}
f^{(3)}&=&[\omega_l+\imath \mu \bar{B}+\hat{k}\vec{\partial}_{r}^{(0)} ]^{-1} D \mbox{,} 
\label{eq:f3shnot}\\
D&=&-\frac{1}{2}\Delta f^{(1)}f^{+(1)} -Pf^{(1)} \label{eq:d3sg3ot}
\mbox{.}
\end{eqnarray}
The first and second term in Eq.~(\ref{eq:d3sg3ot}) 
gives $f^{(3)}_{N}$and $f^{(3)}_{M}$ respectively. The term $P$ is of 
order $\epsilon^2$ and is given by   
\begin{equation}
\label{eq:ttpoeps2}
P=\imath \frac{\mu}{\bar{B}}\bar{\vec{B}}\vec{B}_{1}^{(2)}-
\imath\hat{k}\vec{A}_1^{(2)}
\mbox{.}
\end{equation}
The magnetic contributions $\vec{B}_{1}=\vec{B}_{1}^{(2)}$ and 
$\vec{A}_1^{(2)}$ must be determined by solving Maxwell's 
equation~(\ref{eq:scvecpot}). Expanding~(\ref{eq:scvecpot})   
one obtains two decoupled equations 
\begin{eqnarray}
& & \left(1-\frac{\mu^2 }{\tilde{\kappa}^2} \right)\vec{B}_{1\parallel}=
-\eta^{0} \bar{\vec{B}}_{\parallel} \label{eq:fobpar1} \mbox{,} 
\label{eq:tefdcob} \\
& & \vec{\nabla}_r \times \left[ 
\left(1-\frac{\mu^2 }{\tilde{\kappa}^2} \right) \vec{B}_{1\perp }
+ \eta^{0} \bar{\vec{B}}_{\perp}
\right]=\vec{\beta} \label{eq:sectbsen} 
\mbox{,}
\end{eqnarray}
for the parallel and perpendicular component 
$\vec{B}_{1\parallel}$ and $\vec{B}_{1\perp }$ of $\vec{B}_{1}$. The quantities
$\eta^{0},\,\vec{\beta}$, which are both of order $\epsilon^2$, are given by 
\begin{equation}
\label{eq:hdw2ki}
\eta^{0} = \frac{t}{\bar{B}} \frac{2\mu}{\tilde{\kappa}^2} \sum_{l=0}^{N_D} 
\overline{\Im g}\mbox{,}\hspace{0.3cm}
\vec{\beta} = \frac{2t}{\tilde{\kappa}^2} \sum_{l=0}^{N_D} 
\overline{\hat{k} \Im g}
\mbox{.}
\end{equation}
The Green's function $g$ in Eq.~(\ref{eq:hdw2ki}) may be 
replaced by $-f^{(1)}f^{+(1)}/2$ under the $\hat{k}$-integral. 
Thus, $\eta^{0},\,\vec{\beta}$ may be calculated by using the 
first order solutions $f^{(1)},\,f^{+(1)}$ as given by 
Eq.~(\ref{eq:fsffotp}).  The solution of 
Eqs.~(\ref{eq:fobpar1}, \ref{eq:sectbsen}) is 
obtained by expanding the unknown variables $\vec{B}_{1\parallel}$, 
$\vec{B}_{1\perp }$ and the parameters $\eta^{0},\,\vec{\beta}$, 
which are all invariant under lattice 
translations, in Fourier series; the corresponding 
Fourier coefficients are denoted by 
$(\vec{B}_{1\parallel})_{l,m}=(B_{1\parallel})_{l,m}\vec{e}_y$, 
$(\vec{B}_{1\perp })_{l,m}=(B_{1\perp })_{l,m}\vec{e}_{z}$ and
$(\eta^{0})_{l,m},\,\vec{\beta}_{l,m}$. The explicit solutions will be 
reported at the end of this subsection.

Given the second order contribution $\vec{B}_1$, the first 
term $G^{(4)}_{c}$ of $f^{(3)}_{M}$ [see Eq.~(\ref{eq:g4mpmp})] can be 
evaluated. To calculate the second term $G^{(4)}_{d}$ one needs, 
in addition, the correction term $\vec{A}_1$ [see 
Eqs.~(\ref{eq:f3shnot})-\ref{eq:ttpoeps2})]. Writing 
$\vec{A}_1=\vec{A}_{1\parallel}+\vec{A}_{1\perp}$, the Fourier coefficients 
of $\vec{A}_{1\parallel},\,\vec{A}_{1\perp}$ may be expressed\cite{KLEIN1} in 
terms of the Fourier coefficients of the induction,
\begin{equation}
\label{eq:acoefitob}
\begin{split}
& (\vec{A}_{1\parallel})_{l,m}=\frac{\imath}{\vec{Q}_{l,m}^{\,2}}Q_{l,m,x} (B_{1\parallel})_{l,m}\vec{e}_z 
\mbox{,}  \\
& (\vec{A}_{1\perp})_{l,m}=\frac{\imath}{\vec{Q}_{l,m}^{\,2}}
(B_{1\perp})_{l,m}  \left( Q_{l,m,y} \vec{e}_x-Q_{l,m,x} \vec{e}_y\right)
\mbox{,}
\end{split}
\end{equation}
using the gauge conditions defined by Eq.~(\ref{eq:gcfa1}). 
The quantities $Q_{l,m,x},\,Q_{l,m,y}$ in Eq.~(\ref{eq:acoefitob}) 
are the $x$ and $y$ components of the reciprocal lattice 
vector $\vec{Q}_{l,m}$ defined in appendix~\ref{sec:gapcorfun}.

Each one of the four terms of order $\epsilon^4$ in 
Eqs.~(\ref{eq:nmpofgs},~\ref{eq:mapofgs}) may be represented 
as a multiple integral and Matsubara sum over the product of 
\emph{two} gap correlation functions. 
What remains to be done is to perform analytically as many 
integrations as possible. The details of the calculation 
will be reported here for the first term $G_{a}^{(4)}$, defined by 
Eq.~(\ref{eq:g4nfirst}); the evaluation of the other three 
terms is similar.   

Using the first order Green's functions~(\ref{eq:fsffotp})
and the definition of the gap correlation 
function~(\ref{eq:defgapcorfun}), the term $G_{a}^{(4)}$ takes 
the form
\begin{widetext}
\begin{equation}
\label{eq:g4nttf2}
G_{a}^{(4)}=-\langle\frac{t}{8} \sum_{l=0}^{N_D} 
\left[
\int_0^{\infty}\mathrm{d}s \int_0^{\infty}\mathrm{d}s_1 \int_0^{\infty}\mathrm{d}s_2\,
\mathrm{e}^{-\omega_s(s+s_1+s_2)} \overline{ V(\vec{r}-s_1\hat{k},\vec{r}+s_2\hat{k})
\left[
V(\vec{r}-s\hat{k},\vec{r})+V(\vec{r},\vec{r}+s\hat{k})
 \right]}+\hspace{0.2cm} c.c.
\right] \rangle
\mbox{.}    
\end{equation}
Expanding $V$  in a Fourier series, the spatial average in 
Eq.~(\ref{eq:g4nttf2}) may be performed and $G_{a}^{(4)}$ takes the 
form  
\begin{equation*}
\label{eq:g4ghtf3}
\begin{split}
G_{a}^{(4)}=&-\frac{t}{2} \sum_{l=0}^{N_D} 
\frac{1}{2\pi}\int_0^{2\pi } \textrm{d}\varphi \,
\sum_{l,m} 
\int_0^{\infty}\mathrm{d}s \int_0^{\infty}\mathrm{d}s_1 \int_0^{\infty}\mathrm{d}s_2\;
\mathrm{e}^{-\omega_l(s+s_1+s_2)}   
V_{l,m}\left(-(s_1+s_2)\hat{k}\right)
V_{l,m}^{\star }\left(s\hat{k}\right)
\cos \left(\vec{Q}_{l,m}\frac{s}{2}\hat{k} \right)\cdot \\
& \left[
\cos \left( \vec{Q}_{l,m} \frac{-s_1+s_2}{2}\hat{k} \right) 
\cos \left(\mu\bar{B}\left(s+s_1+s_2 \right) \right)+
\sin \left( \vec{Q}_{l,m} \frac{-s_1+s_2}{2}\hat{k} \right) 
\sin \left(\mu\bar{B}\left(s+s_1+s_2 \right) \right)
\right]
\mbox{.}    
\end{split}
\end{equation*}
where $V_{l,m}$ is given by Eq.~(\ref{eq:frffspa}) and the 
symmetry relations~(\ref{eq:gensymrel}) have been used
to rearrange the integrand. We introduce center of mass
coordinates $t_S=s_1+s_2$, $t_R=s_1-s_2$ in the $s_1,s_2$-plane. 
Replacing $s_1,s_2$ by the new variables, the integration 
over $t_R$ may be performed and the double 
integral over $s$ and $t_S$ becomes a  
product of two independent, one-dimensional integrals. 
Performing this step, $G_{a}^{(4)}$ takes the form      
\begin{equation}
\label{eq:g4gwtf4}
\begin{split}
G_{a}^{(4)}=&-t \sum_{l=0}^{N_D} 
\frac{1}{2\pi}\int_0^{2\pi } \textrm{d}\varphi \,
\sum_{l,m} \frac{1}{\vec{Q}_{l,m}\hat{k}}
\int_0^{\infty}\mathrm{d}s \;
\mathrm{e}^{-\omega_ls}
\cos \left(\frac{s}{2}\vec{Q}_{l,m}\hat{k} \right)\cdot
V_{l,m}^{\star}\left(s\hat{k}\right)
\cdot \\
& \int_0^{\infty}\mathrm{d}t_S \;
\mathrm{e}^{-\omega_lt_S}
\sin \left(\frac{t_S}{2}\vec{Q}_{l,m}\hat{k} \right)\cdot
V_{l,m}\left(-t_S\hat{k}\right)
\left[
\cos \left(\mu\bar{B}s\right)
\cos \left(\mu\bar{B}t_S\right)-
\sin \left(\mu\bar{B}s\right)
\sin \left(\mu\bar{B}t_S\right)
\right]
\mbox{.}    
\end{split}
\end{equation}
\end{widetext}
An attempt to simplify Eq.~(\ref{eq:g4gwtf4}) further, 
by performing one of the remaining integrations 
analytically, was not successful. At this point it seems 
already feasible to calculate the remaining integrals over 
$s,\varphi$ and the sums over Matsubara and Fourier indices 
numerically. However, we prefer to proceed and calculate 
the remaining integrals by means of an asymptotic 
approximation.
 
Let us consider for definiteness the integral over $s$ 
in Eq.~(\ref{eq:g4gwtf4}). The integrand  has its maximum at
$s=0$. We analyze the behavior of the various factors in  
the integrand as a function of $s$, and neglect the 
$s-$dependence of the slowest varying factors.
The characteristic lengths in $s-$space of
the factors $\exp(-\omega_ls),\;
\cos \left(s\vec{Q}_{l,m}\hat{k} \right),\;
V_{l,m}(s\hat{k})$ and 
$\cos \left(\mu\bar{B}s\right)$
are given by 
$\tau_1=[(2l+1)t]^{-1},\;\tau_2=(\bar{B}_{\perp} |\vec{r}_{-l,m}|)^{-1},
\;\tau_3=(n\bar{B}_{\perp})^{-1/2}$ and 
$\tau_4= (\mu\bar{B})^{-1}$, where 
$\vec{r}_{lm}=l\vec{a}+m\vec{b}$. We consider a range of 
inductions $\bar{B} \lesssim B_P$, where the Pauli critical
field $B_P$ is (in the present system of units) given by
$\mu B_P \doteq 0.4$. As a consequence $\tau_4 \gtrsim 2$. Choosing a typical 
number $\mu = 0.1$ for the dimensionless magnetic moment, 
our induction varies in the range $\bar{B} \lesssim 4$. The 
characteristic lengths $\tau_2$ and $\tau_3$ both depend on the Landau
quantum number $n$; recall that $\bar{B}_{\perp}$ depends on $n$
as shown in Figure~(\ref{fig:phabou}). Let us consider 
first the case $n=1$. Then, $\bar{B}_{\perp}\cong\bar{B}\sin\Theta_1$ with
$\sin\Theta_1\approx\mu/\pi$ according to Eq.~(\ref{eq:docan1}) and the 
definition of $\mu$ in appendix~\ref{sec:sou}. As a 
consequence, $\tau_{3}$ is of the same magnitude as $\tau_4$ for 
$n=1$. The magnitude of of $\tau_2\cong7/|\vec{r}_{l,m}|$ varies strongly
depending on the Fourier indices $l,m$. For not too large 
Fourier indices and nearly all $\omega_l$, $\tau_1$ will be the 
smallest of the four characteristic length. This is, 
however, only true for not too low temperatures $t$. 
For large Fourier indices, which should be taken into 
account in the present situation, the behavior of the 
integrand will be dominated by the term 
$\cos \left(s\vec{Q}_{l,m}\hat{k} \right)$  because its 
characteristic length $\tau_2$ becomes small for large $l,m$. 
Thus, the latter term as well as the Matsubara term 
$\exp(-\omega_ls)$ has to be kept, while the terms 
$V_{l,m}(s\hat{k})$ and $\cos \left(\mu\bar{B}s\right)$ show 
the slowest variation in $s$ and may be replaced 
by their values at $s=0$. This conclusion remains true 
for arbitrary $n$. This may be seen by using the relation   
$n\bar{B}_{\perp}\cong\beta$ which will be derived in 
section~\ref{sec:tttppr}. 

Using this asymptotic approximation both the integral
over $s$ and the Fermi surface average may be performed
analytically and one arrives at the result
\begin{equation*}
\begin{split}
G_{a}^{(4)}=&-\frac{t}{4} \sum_{l=0}^{N_D} \sum_{l,m} 
V_{l,m}^{\star}\left(s\hat{k}\right)\big|_{s=0}
V_{l,m}\left(-t_S\hat{k}\right)\big|_{t_s=0}\cdot\\
& \frac{2\omega_l^2+\frac{1}{4}|\vec{Q}_{l,m}|^2}
{\omega_l^2\left(\omega_l^2+\frac{1}{4}|\vec{Q}_{l,m}|^2 \right)^{\frac{3}{2}}}
\mbox{.}\label{eq:g4af2h}    
\end{split}
\end{equation*} 
The second nonmagnetic term $G_{b}^{(4)}$ [see 
Eq.~(\ref{eq:g4nsecond})], which is evaluated with the 
same method, is given by $G_{b}^{(4)}=-2G_{a}^{(4)}$.

In order to calculate the fourth order terms of magnetic 
origin, $G_{c}^{(4)}$ and $G_{d}^{(4)}$, the Fourier coefficients  of 
the quantities $\eta^{0},\,\vec{\beta}$ [see Eq.~(\ref{eq:hdw2ki})] 
have to be evaluated first. This may be done using a 
method similar to the one outlined above for $G_{a}^{(4)}$. A 
noticeable difference is, that the $s-$dependence of the 
slowest varying factors (the ones with characteristic 
lengths $\tau_3$ and $\tau_4$) cannot be completely neglected in 
the course of the asymptotic approximation, but must be 
taken into account to linear order in $s$. In a second 
step, Maxwell's equation has to be solved to obtain 
the magnetic field correction $\vec{B}_1$. Given the 
latter, the free energy $G_{c}^{(4)}$ may be calculated. The 
term $G_{d}^{(4)}$ contains an additional $s-$integral which 
may be performed by means of an asymptotic approximation 
of the above type. The relation $G_{d}^{(4)}=-2G_{c}^{(4)}$ was again 
found to be true, in analogy to the nonmagnetic case 
(the same relation has been found in microscopic 
calculations\cite{RAMPESCH} of the ordinary vortex 
lattice near $H_{c2}$). 

Collecting all fourth order terms and attaching the factor
$\epsilon^4$ one obtains the final result
\begin{widetext}
\begin{equation}
  \label{eq:g4fin1}
G^{(4)}=
\epsilon^4 \Bigg(
\frac{t}{4} \sum_{l,m} f_1^2(x_{l,m}) S_{l,m}^{(1)} 
-\frac{t^2}{\tilde{\kappa}^2-\mu^2} \sum_{l,m}^{\prime}   
\Big[ 
\bar{B}^{\,2}_{\parallel} \mu^4  f_1^2(x_{l,m}) (S_{l,m}^{(1)})^2 +
\left(
\bar{B}_{\perp } \mu^2  f_1(x_{l,m}) S_{l,m}^{(1)} - g_1(x_{l,m}) S_{l,m}^{(2)}
 \right)^2
\Big]
\Bigg)
\mbox{,} 
\end{equation}
\end{widetext}
where the prime at the second summation sign indicates
that the term $l=0,\,m=0$ is to be excluded from the sum.
The functions $f_1,\,g_1$ depend explicitly on the Landau 
quantum number $n$ and are given by 
\begin{eqnarray}
  \label{eq:s2helpf}
f_1(x)&=&\textrm{e}^{-\frac{x}{2}} L_n(x) \\ 
g_1(x)&=&\textrm{e}^{-\frac{x}{2}} 
\left[ 
\frac{1}{2}L_n(x)+\left(1-\delta_{n,0} \right) L_{n-1}^1(x)
\right]
\mbox{,}
\end{eqnarray}
where $L_{n}^1$ is a Laguerre polynomial\cite{GRADRYSH}. 
The Matsubara sums are given by 
\begin{align}
S_{l,m}^{(1)}&= \sum_{l=0}^{N_D} 
\frac
{2 \omega_l^2+\frac{1}{4}|\vec{Q}_{l,m}|^2 }
{
\omega_l^2\left(\omega_l^2+\frac{1}{4}|\vec{Q}_{l,m}|^2  \right)^{3/2}
}\mbox{,} \label{s1matsum}\\  
S_{l,m}^{(2)}&= \sum_{l=0}^{N_D} 
\frac{1}
{
\omega_l^2\left(\omega_l^2+\frac{1}{4}|\vec{Q}_{l,m}|^2  \right)^{1/2}
}\mbox{.}    
\end{align}
The square of the reciprocal lattice vector is conveniently
written in the form $|\vec{Q}_{l,m}|^2=2\bar{B}_\perp x_{l,m}/2$ where  
$x_{l,m}$ is defined by Eq.~(\ref{eq:affs6k}). Introducing a 
magnetic length $L$ defined by  
\begin{equation}
  \label{eq:defofmagnl}
\bar{B}_\perp=\frac{2\pi}{ab\sin\alpha}=\frac{2}{L^2}
\mbox{,}
\end{equation}
these parameters which depend on $a,\,b,\,\alpha$, are 
given by  
\begin{equation}
  \label{eq:xlmterm}
x_{l,m}=\frac{\pi^2}{\sin^2\alpha}
\left(\frac{L}{a}\right)^2 l^2+
\left(\frac{a}{L}\right)^2 m^2-
2 \pi lm\frac{\cos\alpha}{\sin\alpha}
\mbox{.}
\end{equation}   

\subsection{Local induction}
\label{subsec:locinduct} 

The components $B_{1\perp}$ and $B_{1\parallel}$ of the spatially varying 
magnetic field $\vec{B}_1$ are given by Fourier series 
of the form
\begin{equation}
  \label{eq:locmfsup}
B_{1\Delta }(\vec{r})= \sum_{l,m}^{\prime} (B_{1\Delta })_{l,m}\textrm{e}^{\imath \vec{Q}_{l,m}\vec{r}}
\mbox{,}
\end{equation}
where $\Delta=\perp,\parallel$.  The Fourier coefficients are given by
\begin{align}
(B_{1\parallel})_{l,m}=&-\frac{t\langle|\Delta_n|^2\rangle}{\tilde{\kappa}^2-\mu^2}
\bar{B}_{\parallel} \mu^2 \cdot  \nonumber \\
&(-1)^{lm} \textrm{e}^{-\imath \pi l \frac{b}{a}\cos\alpha } f_1(x_{l,m}) S_{l,m}^{(1)} \mbox{,}
\label{eq:b1par}  \\
(B_{1\perp})_{l,m}=& 
-\frac{t\langle|\Delta_n|^2\rangle}{\tilde{\kappa}^2-\mu^2}
(-1)^{lm} \textrm{e}^{-\imath \pi l \frac{b}{a}\cos\alpha }  \cdot  \nonumber \\ 
&\left(\bar{B}_{\perp } \mu^2  f_1(x_{l,m}) S_{l,m}^{(1)} - g_1(x_{l,m}) S_{l,m}^{(2)}\right)  
\label{eq:b1senk}  
\mbox{.}
\end{align}
The parallel component~(\ref{eq:b1par}) is proportional 
to $\mu^2$ and is entirely due to the spin pair-breaking 
effect. The perpendicular component~(\ref{eq:b1senk}) is 
the sum of a $\mu^2$-dependent term and a second term not 
(explicitly) dependent on $\mu$. The terms dependent 
on $\mu^2$ have the same form for both components (recall 
that the direction of $B$ in spin space is arbitrary) 
and are proportional to the relevant component of the 
macroscopic induction. The second term in 
Eq.~(\ref{eq:b1senk}), which is of opposite sign, may 
only for $n=0$ be considered as a consequence of 
orbital pair-breaking; for $n>0$ this second term 
depends also (since a positive $n$ is necessarily due 
to a finite $\mu$) on the spin pair-breaking effect. The 
GL limit of the local induction is discussed in 
appendix~\ref{sec:tgll}.      

The validity of the asymptotic approximation 
used in the derivation of 
Eqs.~(\ref{eq:g4fin1},\ref{eq:locmfsup}) is 
not restricted to low $n$, but sufficiently high 
temperatures, say $t>0.1$, should be used. Clearly, 
if different states with \emph{very small} free energy 
differences are found, no conclusion as to the 
relative stability of these states can be drawn.

\subsection{Extremal conditions}
\label{subsec:extrcond} 
In thermodynamic equilibrium, the values of 
$\epsilon,\,\bar{B}_\perp,\,\bar{B}_{\parallel}$ and the lattice parameters 
$a,\,b,\,\alpha$ have to be chosen in such a way that the 
free energy becomes minimal. To find the equilibrium 
values of $\epsilon,\,\bar{B}_\perp,\,\bar{B}_{\parallel}$ the extremal conditions
\begin{equation}
\label{eq:extrcfa3}
\frac{\partial G}{\partial\epsilon}=0,\:\frac{\partial G}{\partial\bar{B}_{\parallel}}=0,\:\frac{\partial G}{\partial\bar{B}_\perp}=0
\mbox{,} 
\end{equation}
have to be solved near $H_{c2}$ . The question for 
the optimal $a,\,b,\,\alpha$ will be addressed in 
the next section.

Inserting the superconducting solution for $\epsilon$ in the 
free energy yields
\begin{equation}
  \label{eq:freengi}
G=\bar{G}-\frac{1}{4}\frac{\left(\bar{G}^{(2)} \right)^2}{\bar{G}^{(4)}}
\mbox{,}
\end{equation}
where the coefficients $\bar{G}\;\bar{G}^{2},\;\bar{G}^{4}$
are defined by $G=\bar{G}+\epsilon^{2}\bar{G}^{(2)}+\epsilon^{4}\bar{G}^{(4)}$. 
Eq.~(\ref{eq:freengi}) shows, that the stable lattice 
structure (see section~\ref{sec:rffpf}) is determined 
by the requirement of minimal $\bar{G}^{4}$

To find the two-component macroscopic magnetization 
relation between induction $\bar{B}_\perp,\bar{B}_{\parallel}$ and 
external field $H_\perp,H_{\parallel}$, the above extremal
conditions must be solved for $\bar{B}_\perp,\;\bar{B}_{\parallel}$. This
cannot be done for arbitrary fields but requires an 
appropriate expansion of the coefficients for small
$\bar{B}_\perp-\bar{B}_{c2,\perp},\;\bar{B}_{\parallel}-\bar{B}_{c2,\parallel}$. A lengthy but
straightforward calculation, generalizing Abrikosovs 
classical work\cite{ABRI1} to the present situation, 
leads to the result 
\begin{equation}
\begin{split}
\label{eq:marebh}
\bar{B}_\perp&=\alpha_{\perp \perp }H_\perp+\alpha_{\perp\parallel }H_{\parallel}+\beta_{\perp}   \\
\bar{B}_{\parallel}&=\alpha_{\perp\parallel  }H_\perp+\alpha_{\parallel \parallel }H_{\parallel}+\beta_{\parallel }
\mbox{.}
\end{split}
\end{equation}
The coefficients in this linear relation are given by
\begin{align}
&\alpha_{\perp \perp }=2\tilde{\kappa}^2\left[2\left(\tilde{\kappa}^2-
\mu^2 \right)-A_{\parallel} \right]/\textrm{det}M  \nonumber \\
&\alpha_{\perp\parallel }=2\tilde{\kappa}^2 A_{\parallel \perp }/\textrm{det}M  \nonumber \\
&\alpha_{\parallel \parallel }= 2\tilde{\kappa}^2\left[2\left(\tilde{\kappa}^2-
\mu^2 \right)-A_{\perp} \right]/\textrm{det}M   \nonumber \\
&\beta_{\perp}=-2\left(\tilde{\kappa}^2-\mu^2 \right)
\left(A_{\perp} B_{c2,\perp}+A_{\parallel\perp }B_{c2,\parallel} \right)/\textrm{det}M  \nonumber  \\
&\beta_{\parallel }=-2\left(\tilde{\kappa}^2-\mu^2 \right)
\left(A_{\parallel } B_{c2,\parallel}+A_{\parallel \perp }B_{c2,\perp} \right)/\textrm{det}M \mbox{,} \nonumber \\
&\mbox{where:}\;  \textrm{det}M=2\left(\tilde{\kappa}^2-\mu^2 \right)
\left[2\left(\tilde{\kappa}^2-\mu^2 \right)-A_{\parallel }-A_{\perp} \right]
\mbox{.}  \nonumber 
\end{align}  
The parameters $A_{\parallel},\ldots$ may be calculated for a given 
lattice structure with the help of the relations 
\begin{align}
A_{\parallel} &=\frac{1}{2\bar{G}^{(4)}}
\left(\frac{\partial\bar{G}^{(2)}}{\partial\bar{B}_{\parallel}} \right)^2   \\
A_{\perp} &=\frac{1}{2\bar{G}^{(4)}} 
\left(\frac{\partial\bar{G}^{(2)}}{\partial\bar{B}_{\perp }} \right)^2   \label{eq:scas3}  \\
A_{\parallel \perp } &=\frac{1}{2\bar{G}^{(4)}}
\frac{\partial\bar{G}^{(2)}}{\partial\bar{B}_{\parallel}}
\frac{\partial\bar{G}^{(2)}}{\partial\bar{B}_{\perp}} \mbox{,}   
\end{align}
where the derivatives of $\bar{G}^{(2)}$ have to be evaluated 
at $\bar{B}=B_{c2}$ and the relation $A_{\parallel}A_{\perp}=A_{\parallel \perp }^2$ may be shown
to be true.

Eq.~(\ref{eq:marebh}) constitutes the macroscopic relation
between induction and external field for a 2D
superconductor in a tilted magnetic field. It is, of course, 
strongly anisotropic and shows a coupling between the 
parallel and perpendicular field components. 
For $H_{\parallel}=0,\,\mu\Rightarrow0$ and $t\Rightarrow1$, Eq.~(\ref{eq:marebh}) should 
reduce to Abrikosov's GL solution\cite{ABRI1,SAJATH} for 
the magnetization of a triangular vortex lattice. This 
is indeed the case as shown in Appendix~\ref{sec:tgll}.

For $H_{\parallel}=0,\,\mu\Rightarrow0$, Eq.~(\ref{eq:marebh}) describes 
the ordinary vortex lattice (near $H_{c2}$) for arbitrary 
temperatures. A numerical comparison with corresponding 
results by Eilenberger\cite{EILE2} and Rammer and 
Pesch\cite{RAMPESCH} has not been undertaken because 
a different (spherical) Fermi surface has been 
used in these works. However, the limit $H_{\parallel}=0,\,\mu\Rightarrow0$ of 
the present theory will be checked 
in appendix~\ref{sec:tlotovl} by calculating the 
critical value of $\kappa$ separating type II 
from type I superconductivity.
   
\section{\label{sec:rffpf} Results for finite perpendicular 
field}

In this section we determine the stable order parameter 
structures for the paramagnetic vortex states with $1 \leq n \leq 4$ 
in the vicinity of $H_{c2}$. The 
numerical procedure to find the stable states is 
essentially the same as in KRS\cite{KLRAISHI}. First, 
the upper critical field 
$B_{c2}$ and the corresponding quantum number $n$ have to be
found for given temperature $t$ and tilt angle $\Theta$
by solving the linearized gap equation~(\ref{eq:g2finres}).
In a second step, the stable lattice 
structure, which minimizes the fourth order term 
$\bar{G}^{(4)}=G^{(4)}/\epsilon^4$ [see Eq.~(\ref{eq:g4fin1}], has to be 
determined. Because of the flux quantization condition 
the minimum with respect to only \emph{two} parameters, 
which may be chosen as $a/L$ and $\alpha$, must be found. 
In contrast to the ordinary vortex lattice, where it is 
usually sufficient to calculate only a few lattices 
of high symmetry (triangular, quadratic) to find the 
stable state, the present situation is characterized 
by a large number of local minima of Eq.~(\ref{eq:g4fin1}),
corresponding to a large number of possible lattices of 
rather irregular shape. Therefore, a graphical method was 
used to determine the stable state; the free energy 
surface $\bar{G}^{(4)}(a/L,\alpha)$ was plotted for the whole 
$(a/L,\alpha)$-plane and the global minimum was determined 
by inspection. Basically, two material parameters, 
$\mu$ and $\tilde{\kappa}$, and two externally controlled 
parameters, $t$ and $\Theta$, enter the theory. Numerical 
calculations have been performed for a single value of 
$\mu=0.1$, two different reduced temperatures $0.2$ and
$0.5$, four different values $0.1,\,1.0,\,10,\,100$ of 
Eilenbergers parameter $\tilde{\kappa}$, and several values of
$\Theta$ corresponding to different Landau quantum numbers 
$n$. Some of the resulting order parameter and magnetic 
field structures in the range $n\leq4$ will be reported here. 
These low-$n$ pairing states are, of course, the most 
important ones from an experimental point of view.

For comparison we consider first, in 
appendix~\ref{sec:tlotovl}, the ordinary vortex 
lattice state with $n=0$. This illustrates the method and
may also be used to check the accuracy of our asymptotic 
approximation. The equilibrium state for low-$\kappa$ type II 
superconductors is calculated and good agreement with 
previous theories is found for not too low temperatures.   

\begin{figure}[htbp] 
\begin{center}\leavevmode
\includegraphics[width=8cm]{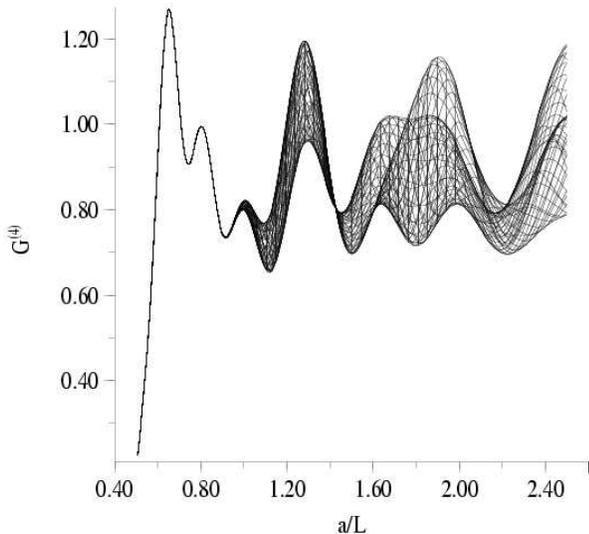}
\caption{\label{fig:g4nproj} (Reduced image quality due to
arXiv restrictions) Projection of the free 
energy $\bar{G}^{(4)}$ on the $\bar{G}^{(4)},a/L-$plane. Using 
this perspective the $\alpha-$independent parts of the free 
energy surface are displayed as lines. In the considered 
range of $a/L$ one finds six local minima, corresponding 
to two FFLO-like and four two-dimensional lattices. The 
global minimum is at $a/L\approx1.1$ and corresponds to a 
two-dimensional lattice. Parameters chosen in this plot 
are $n=7,\;t=0.5,\;\tilde{\kappa}=10,\;\mu=0.1,\;\Theta=0.055$.}
\end{center} 
\end{figure}

Considering now pairing states with $n>0$, the number of 
order parameter zeros per unit cell increases clearly 
with increasing $n$. One 
finds\cite{KLRAISHI} two types of minima of 
$\bar{G}^{(4)}(a/L,\alpha)$, isolated minima and line-like minima. 
The first type corresponds to 
``ordinary'' 2D lattices, the second type, 
characterized in a contour plot [see Fig.1 of 
KRS~\cite{KLRAISHI}] by a line of constant 
$a/L$ with $\bar{G}^{(4)}(a/L,\alpha)$ nearly independent of $\alpha$,  
corresponds to quasi-one-dimensional, or ``FFLO-like'' 
lattices (rows of vortices and one-dimensional FFLO-like 
minima alternating). A convenient way to identify 
the type of minimum and find
its position on the $a/L-$axis, is to plot the 
projection of the $\bar{G}^{(4)}(a/L,\alpha)-$surface on 
the $(\bar{G}^{(4)},a/L)-$plane. An example for this 
perspective, where $\alpha-$ independent parts of the 
free energy surface show up as lines, is given 
in Fig~\ref{fig:g4nproj} for $n=7$. The $\alpha-$coordinate of 
a 2D minimum cannot be read off from 
such a plot and requires a second projection on 
the $(a/L,\alpha)-$plane (such as Fig~\ref{fig:ordvlatt}
or Fig.1 of KRS\cite{KLRAISHI}). The free energy maps 
for other $n>0$ states are in principle similar to 
Fig~\ref{fig:g4nproj} but the different local minima 
show more pronounced differences for smaller $n$ . 
 
\begin{figure}[htbp] 
\begin{center}\leavevmode
\includegraphics[width=8cm]{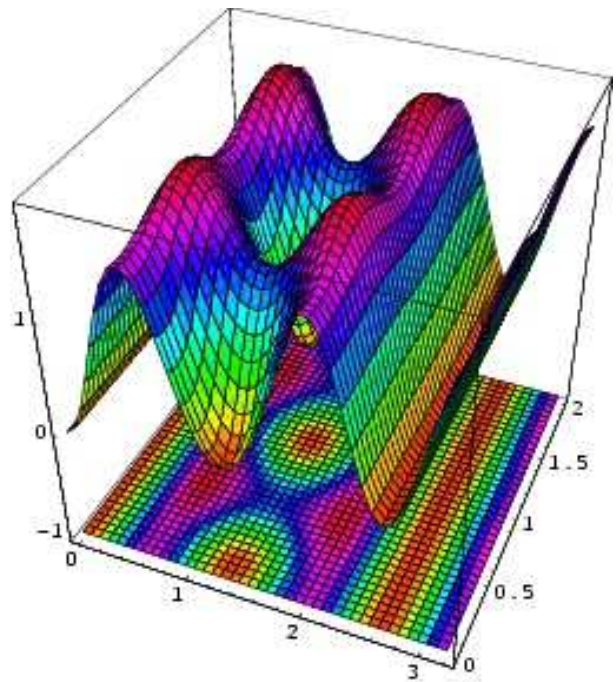}
\caption{\label{fig:opn1k100} (Reduced image quality due to
arXiv restrictions) Square of 
modulus of order parameter $|\psi_1|^2$ as a function of $x/a,\,y/a$
in the range $0<x/a<2,\;0<y/a<3.2$. This is the 
stable structure (unit cell parameters 
$a/b=0.205,\,\alpha=33\mbox{{\r{}}}$) for 
$t=0.2,\,\tilde{\kappa}=100,\,\theta=1.2\mbox{{\r{}}}\,
(n=1,\,B_{c2}=4.141)$}.
\end{center} 
\end{figure}

Let us start with the paramagnetic vortex state 
with n=1 and consider first the limit 
of large $\kappa$. As reported in KRS\cite{KLRAISHI}, a 
quasi-one-dimensional state is found to 
be stable in this case. Fig~\ref{fig:opn1k100} 
shows the spatial variation of the modulus of the order 
parameter. One sees rows of vortices separated by a single,
FFLO-like line of vanishing order parameter. The unit
cell of the structure shown in Fig~\ref{fig:opn1k100} is
given by $a/L=1.0875,\,\alpha=33\deg$. A shift of the vortex rows 
relative to each other leads to a lattice with the same
$a/L$ and a different $\alpha$, which has nearly the same 
free energy (which is reasonable, since the interaction
between vortices from different rows is weak as a 
consequence of the intervening FFLO domain wall).
The vortices are of the ``ordinary'' type, i.e. the 
phase of the order parameter changes by $+2\pi$ when 
surrounding the center.

\begin{figure}[htbp]  
\begin{center}\leavevmode
\includegraphics[width=8cm]{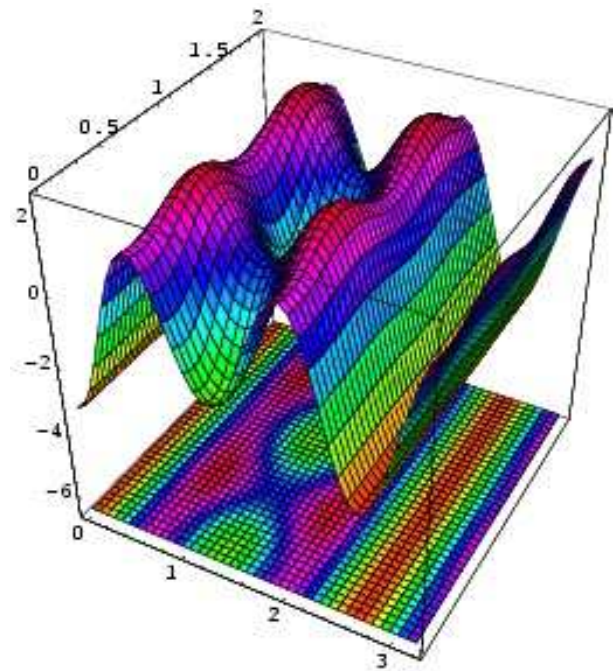}
\caption{\label{fig:b1parn1k100} (Reduced image 
quality due to arXiv restrictions)  
Parallel component $B_{1\parallel}$ as a function of $x/a,\,y/a$ 
in the range $0<x/a<2,\;0<y/a<3.2$. This plot has been 
produced using the same input parameters as in 
Fig~\ref{fig:opn1k100}.}
\end{center} 
\end{figure}

It is of interest, to calculate the magnetic field 
belonging to this order parameter structure. We plot 
the parallel and perpendicular components $B_{1\parallel }(\vec{r})$ 
and $B_{1\perp }(\vec{r})$ of the spatial varying part 
$\vec{B}_1(\vec{r})$ of the magnetic field as given by 
Eqs.~(\ref{eq:locmfsup})-(\ref{eq:b1senk}),
omitting a common factor $t\langle|\Delta_n|^2\rangle/(\tilde{\kappa}^2-\mu^2)$. The 
field $B_{1\parallel }(\vec{r})$, which is entirely due 
to the spin pair-breaking mechanism, is shown in 
Fig~\ref{fig:b1parn1k100}. Due to its paramagnetic nature,
the field $B_{1\parallel }(\vec{r})$ is \emph{expelled} from regions of 
small $\psi(\vec{r})$. This behavior is exactly opposite 
to the usual orbital response, which implies an enhancement 
of the induction in regions of small $|\psi|(\vec{r})$. As 
a consequence, the spatial variation of $B_{1\parallel }$ is very similar
to that of $|\psi|^2$, shown in Fig~\ref{fig:opn1k100}.

\begin{figure}[htbp]  
\begin{center}\leavevmode
\includegraphics[width=8cm]{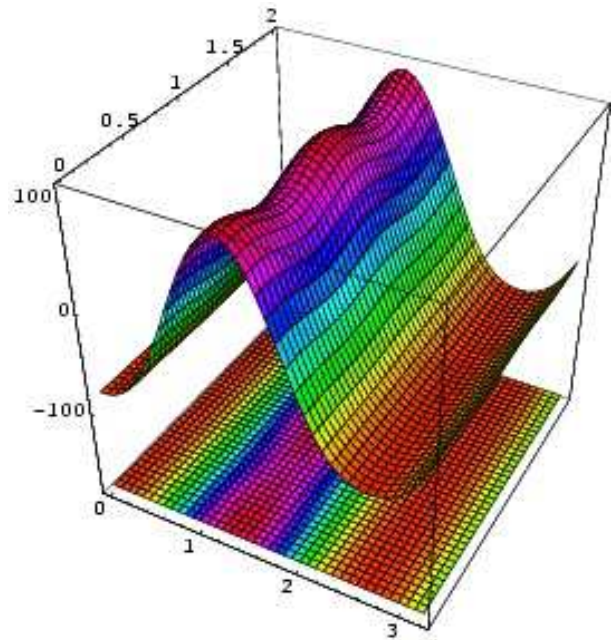}
\caption{\label{fig:b1perpn1k100} (Reduced image 
quality due to arXiv restrictions)   
Perpendicular component $B_{1\perp }$ as a function of 
$x/a,\,y/a$ in the range $0<x/a<2,\;0<y/a<3.2$. The 
same input parameters as in Fig~\ref{fig:opn1k100} 
have been used.}
\end{center} 
\end{figure}

The perpendicular  field $B_{1\perp }(\vec{r})$, shown in 
Fig~\ref{fig:b1perpn1k100}, consists of a spin term  
proportional to $\mu^2$, and a second term which depends 
[see Eq.~(\ref{eq:b1senk})] not explicitly on $\mu$. 
The term proportional to $\mu^2$ is negligibly small and 
the total field is essentially given by the second term. 
Near the vortices the field $B_{1\perp }(\vec{r})$ behaves 
in the familiar, orbital way, i.e. it is largest at the 
points of vanishing $\psi$ and decreases with increasing 
distance from the vortex centers. However, at the FFLO-like 
lines of vanishing order parameter, where no topological 
singularity occurs, $B_{1\perp }$ has a \emph{minimum}, i.e. shows 
paramagnetic behavior. Thus, the magnetic response of 
a $n=1$ superconductor may either lead to a local 
suppression or to an enhancement of the magnetic field 
in regions of small order parameter. This is in contrast 
to the purely orbital response of a $n=0$ superconductor, 
where the magnetic field is always enhanced.   
This unconventional behavior is formally due to the second 
term in $g_1$ [see Eq.~(\ref{eq:s2helpf})]. The field 
$B_{1\parallel }$ is much smaller than $B_{1\perp }$ and the total field 
$B_1$ for $n=1$ is consequently dominated by the 
perpendicular component $B_{1\perp }$, which is a consequence 
of the combined action of both pair-breaking mechanisms.

The quasi-one-dimensional order parameter structure shown 
in Fig~\ref{fig:opn1k100} seems to be representative for 
the pairing state with $n=1$; no other stable  
state has been found for $\tilde{\kappa}=10$ and $t=0.5$.
At $\tilde{\kappa}=1,\,0.1$ the free energy surface has no 
minimum at all, which means that a transition to type I 
superconductivity occurs at some value of $\tilde{\kappa}$ 
between $1$ and $10$.       

Extrapolating the $n=1$ result to higher $n$, one would 
expect the following structure for the pairing state with 
Landau quantum number $n$: rows of vortices separated by 
$n$ lines of vanishing order parameter . Such a structure 
would approach the (line-like) FFLO state in the limit $n\Rightarrow\infty$.
However, this simple picture is not realized, at least in 
the important range of low $n$. It holds generally 
for odd $n$, but for even $n$ \emph{two-dimensional} 
structures are preferred. In the latter case, one 
has $n+1$ isolated order parameter zeros per unit cell, with 
associated phase changes of a multiple of $2\pi$. Such a 
situation leads necessarily to the presence of one or more 
\emph{antivortices} - vortices with a topological phase 
change of $-2\pi$ around the center - for states with even 
$n$, since the total phase change around the unit cell must 
remain $+2\pi$. Recently, various proposals to create stable 
antivortices have been published; see e.g. Moshkalkov et al
\cite{MFDM}. In the present context, it is clearly 
the strong paramagnetic pair-breaking, which is responsible 
for the stability of the antivortices. Among the (even-$n$) 
antivortex states, the one with $n=2$ is most easily 
accessible from an experimental point of view and very 
stable under variations of $t$ and $\kappa$. Its properties will 
be discussed in detail in a separate publication\cite{KLTBP}; 
a preliminary account has been published already\cite{KLEIN4}.

For $n=3$ free energy minima for spatially varying states 
exist in the whole considered range $0.1 \leq \kappa  \leq 100$
of the GL parameter. Thus, increased spin pair-breaking 
stabilizes inhomogeneous equilibrium structures 
and shifts the phase boundary between type II and type I 
superconductivity to lower values of $\kappa$.
In the high-$\kappa$ region (for $\kappa \geq 10$) the 
stable state of a $n=3$ superconductor is of the 
quasi-one-dimensional type (at lower $\kappa$ a 2D 
state of nearly the same free energy has been found, which 
will not be discussed here). 
The fields $|\psi|^2$, $B_{1\parallel}$ look similar to the $n=1$ case 
(see Figs.~\ref{fig:opn1k100} and~\ref{fig:b1parn1k100}) 
except that the vortex rows are now separated by 
\emph{three} FFLO-like lines of vanishing order parameter. 
The vortices in neighboring rows are already completely 
decoupled for $n=3$; a translation of neighboring rows 
relative to each other changes the angle $\alpha$ between 
the unit cell basis vectors but does not lead to any change 
(within 8 digits) of the free energy. The perpendicular 
induction $B_{1\perp }$ is again dominated by the second term in 
Eq.~(\ref{eq:b1senk}) and looks similar to the $n=1$ case 
(see Fig.~\ref{fig:b1perpn1k100}); in contrast to the spin part 
this field does not reflect the detailed order parameter 
structure but has only a single broad minimum at the position 
of the three FFLO lines.

\begin{figure}[htbp]  
\begin{center}\leavevmode
\includegraphics[width=8cm]{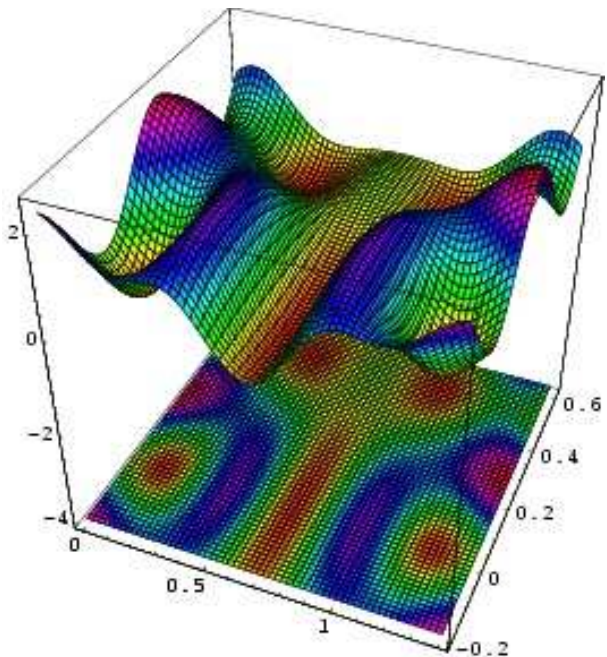}
\caption{\label{fig:opn4k10} (Reduced image 
quality due to arXiv restrictions)   
Square of modulus of order parameter $|\psi_4|^2$ as a 
function of $x/a,\,y/a$ in the range 
$0<x/a<1.4,\;-0.2<y/a<0.62$. This is the 
stable structure (unit cell parameters 
$a/b=0.6735,\,\alpha=70.125\mbox{{\r{}}}$) for 
$t=0.5,\,\tilde{\kappa}=10,\,\theta=0.1\mbox{{\r{}}}\,
(n=4,\,B_{c2}=3.486)$}.
\end{center} 
\end{figure}

For a $n=4$ superconductor at $t=0.5$ the equilibrium state 
is of the quasi-one-dimensional type for $\kappa \leq 1$, and 
of the 2D type for $\kappa \geq 10$. 
Fig.~\ref{fig:opn4k10} shows the 2D order 
parameter structure for a superconductor with $\kappa=10$. There 
are $5$ zeros of $\psi$ per unit cell, one of them of an 
elongated shape. The nature of these topologically singular 
points may be clarified by plotting either the 
phase\cite{KLEIN4} or the local magnetic field.
The parallel component $B_{1\parallel }(\vec{r})$ of the field 
$\vec{B}_1(\vec{r})$ is again 
(compare Figs.~\ref{fig:opn1k100} and~\ref{fig:b1parn1k100}) 
similar in shape to the order 
parameter $|\psi|^2$ and need not be displayed here. 
\begin{figure}[htbp]  
\begin{center}\leavevmode
\includegraphics[width=8cm]{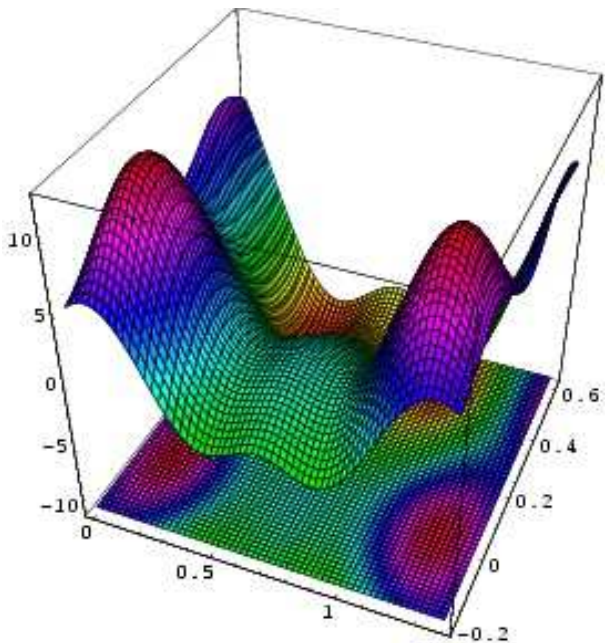}
\caption{\label{fig:b1perpn4k10} (Reduced image 
quality due to arXiv restrictions)  
Perpendicular component $B_{1\perp }$ as a function of 
$x/a,\,y/a$ in the range $0<x/a<1.4,\;-0.2<y/a<0.62$. 
The same input parameters as in Fig~\ref{fig:opn4k10} 
have been used.}
\end{center} 
\end{figure}
The perpendicular field $B_{1\perp }$ is shown in 
Fig.~\ref{fig:b1perpn4k10}. Three of the $5$ order parameter
zeros displayed in Fig~\ref{fig:opn4k10} belong to 
"ordinary" vortices, with local field enhancement and 
diamagnetic screening current (two of the three maxima of
$B_{1\perp }$ are pronounced, while the third, the one corresponding 
to the elongated zero of $\psi$, is rather flat). The 
remaining $2$ order parameter zeros belong to antivortices 
with opposite sign of the "screening currents" (which are 
now paramagnetic in nature) and with minima of $B_{1\perp }$ at the 
points of vanishing $\psi$.  

Results for $n>4$ will not be reported here. Many 
interesting and complex structures may be produced for 
larger $n$. However, the number of different states with 
similar free energies increases with increasing $n$.
As a consequence, the approximate nature of our analytical 
calculation does not allow an identification of the 
stable state for large $n$ . At the same time, an experimental 
verification of these large-$n$ states seems 
difficult since a very precise definition of 
the tilt angle $\theta$ would be required.

\section{\label{sec:sotfs} Structure of the FFLO state}

The stable state in the purely paramagnetic limit $n\to\infty$
has been determined first by Larkin and Ovchinnikov\cite{LAOV} 
at $T=0$ for a spherical Fermi surface. They predicted a 
one-dimensional periodic order parameter structure of the 
form $\Delta(\vec{r})\approx \cos (\vec{q}\vec{r})$, which will be 
referred to as LO state. Later, various 
analytical investigations of the stable states in the vicinity 
of $T_{tri}$ and near $T=0$ have been performed\cite{BUZKUL,MORCOM}
; many other references may be found in a recent review 
article\cite{CASNAR}. A careful search for the state of
lowest free  energy, comparing several possible 
lattices in the whole temperature range, has been reported 
by Shimahara\cite{HIR4}. He found that 
below $t=0.24$ various 2D periodic states
have lower free energy than the one-dimensional 
$\cos (\vec{q}\vec{r})$ state. Shimahara uses the same 
cylindrical Fermi surface as we do and his results do 
therefore apply to the present problem. Nevertheless, we
reconsider in this section the problem of the determination 
of the FFLO structure, in order to have a complete 
description of all states in a tilted field in a single 
theoretical (quasiclassical) framework.
  
The results derived in section ~\ref{sec:entucf}
cannot be used to perform the limit $n\to\infty$ and determine 
the stable state in the purely paramagnetic limit. 
However, the general formalism may be applied in a 
straightforward way to the simpler case of vanishing 
vector potential. In order to be able to compare with 
previously published results we neglect in this section 
the possibility of spatial variations of $\vec{B}$ and 
restrict ourselves to the high-$\kappa$ limit. 

The space of basis functions, which has to be used to expand 
all variables near $H_{FFLO}(T)$, is now given by the infinite 
set $\exp(\imath \vec{q}\vec{r})$ with a fixed value of 
$|\vec{q}|$. Usually, one assumes that the order parameter 
$\Delta$ fulfills some further symmetry (or simplicity) 
requirements, which then leads to a strong decrease of 
the number of unknown coefficients. Following this 
convention, we restrict ourselves to two- and 
one-dimensional periodic structures. For $\Theta>0$, the order 
parameter is not periodic but changes its phase by 
certain factors under translations between equivalent 
lattice points. These phase factors are proportional to 
the perpendicular induction [cf. Eq.~(\ref{eq:defgapcorfun})] 
and vanish for $\Theta\to0$. Thus, the assumption of a periodic 
order parameter for $\Theta=0$ is reasonable (though not 
stringent). It implies, that all allowed wave vectors 
in the expansion of $\Delta$ must be vectors of a reciprocal 
lattice. 

A further slight simplification stems from the behavior of 
the quasiclassical equations under the transformation 
$\vec{r}\Rightarrow-\vec{r},\,\vec{k}\Rightarrow-\vec{k}$, which implies 
that the order parameter must be either even or odd under 
a space inversion $\vec{r}\Rightarrow-\vec{r}$. Thus, the order 
parameter may be written as an infinite sum
\begin{equation}
  \label{eq:ffloopans}
\Delta(\vec{r})= \sum_{m} \Delta_{m}\textrm{e}^{\imath \vec{Q}_{m}\vec{r}}
\mbox{,}
\end{equation}
with coefficients defined by 
\begin{equation}
  \label{eq:ffloocoeff}
\Delta_{m}= |\Delta| \sum_{i=1}^{I} c_{i}\left(\delta_{m,n_{i}} \pm \delta_{m,-n_{i}}  \right)
\mbox{.}
\end{equation}
Here, a shorthand notation $m$ is used for the two 
integers characterizing a 2D reciprocal 
lattice vector $\vec{Q}_{m}$ [cf. the Fourier expansion at the 
beginning of appendix~\ref{sec:gapcorfun}]. The vectors actually 
entering the expansion are distinguished by an 
index $i$, their total number is $I$, and  
the two integers characterizing $\vec{Q}_{n_i}$  
are denoted by $n_i$. The complex numbers $c_i$ are 
the expansion coefficients; one may set $c_1=1$ since 
only the relative weight is important. It turns out, 
that the two solutions distinguished in 
Eq.~(\ref{eq:ffloocoeff}) by a sign are essentially 
equivalent, and only one of them, say the even one, need 
be considered. Thus, the order parameter becomes a 
linear combination of cosine functions.

All reciprocal lattice vectors used in 
Eq.~(\ref{eq:ffloopans}) must be of the same length. 
Denoting this length by $q(T)$, the condition 
$|\vec{Q}_{m}|=q(T)$ takes the form     
\begin{equation}
  \label{eq:condmlatt}
\left(\frac{l}{\tilde{a}} \right)^{2}\frac{1}{\sin^{2}\alpha_{p}}+
\left(\frac{j}{\tilde{b}} \right)^{2}\frac{1}{\sin^{2}\alpha_{p}}-
2\frac{lj}{\tilde{a}\tilde{b}}\frac{\cos \alpha_{p}}{\sin^{2}\alpha_{p}}
=1
\mbox{,}
\end{equation}

where the two integers $l,\,j$  have been used here to 
represent the double index $m$.
The dimensionless quantities $\tilde{a},\,\tilde{b}$ are 
defined by $\tilde{a}=q(T)a_p/2\pi,\,\tilde{b}=q(T)b_p/2\pi$, where  
$a_p,\,b_p,\,\alpha_p$ denote the lattice parameters in the paramagnetic 
limit. If $I$ reciprocal lattice vectors exist, the lattice 
parameters $a_p,\,b_p,\,\alpha_p$, fulfill $I$ relations like 
Eq.~(\ref{eq:condmlatt}) with $I$ pairs of 
integers $l_1,j_1,\ldots l_I,j_I$.
     
Using Eq.~(\ref{eq:ffloopans}) the 
free energy expansion near $H_{FFLO}(T)$, including terms of 
fourth order in the small amplitude $|\Delta|$, may be 
performed by means of methods similar to 
section~\ref{sec:entucf}. The result for the purely 
paramagnetic free energy $G_p$ takes the form   
\begin{equation}
  \label{eq:fenparlim}
G_p=\bar{G}_{p}+G_{p}^{(2)}+G_{p}^{(4)}
\mbox{,}
\end{equation}
where $\bar{G}_{p}=-\mu^2H^2$, and $G_{p}^{(2)}$  and $G_{p}^{(4)}$ are 
contributions of order $|\Delta|^2$ and $|\Delta|^4$ respectively.

The second order term is given by 
\begin{equation}
  \label{eq:fenparsec}
G_{p}^{(2)}=|\Delta|^2 \sum_{i=1}^{I}\,|c_i|^2 \bar{A}
\mbox{,}
\end{equation}
with the $i-$independent coefficient $\bar{A}$ defined by 
\begin{equation*}
\bar{A}= 2 \Big(
\ln t +  t \int_0^{\infty}\textrm{d}s\, \frac{1- \textrm{e}^{-\omega_Ds}} 
{\sinh st}\big[1 - \\
 \cos(\mu \bar{B} s) J_0(sq)
 \big]
\Big)
\mbox{.}
\end{equation*}
The condition $\bar{A}=0$ determines the 
upper critical field; it may also 
be derived from Eq.~(\ref{eq:g2finres}), performing the 
limit $n\to\infty$.

The fourth order term is given by 
\begin{equation}
  \label{eq:fenparfourth}
\begin{split}
G_{p}^{(4)}=|\Delta|^4
\Big[ 
&\sum_{i=1}^{I}\,|c_i|^4 \bar{A}_{i}+ \sum_{i\neq k}^{I}\,|c_i|^2 |c_k|^2 \bar{B}_{i,k}+\\
&\sum_{i\neq k}^{I}\,\left[\left(c_i^{\star} \right)^{2} 
\left(c_k \right)^{2}+c.c.\right] \bar{C}_{i,k}
\Big]
\end{split}
\mbox{,}
\end{equation}
$G_{p}^{(4)}$ depends on the lattice structure via the 
coefficients $\bar{A}_{i}$, $\bar{B}_{i,k}$, and $\bar{C}_{i,k}$, 
which are defined by     
\begin{equation*}
\begin{split}
&\bar{A}_{i}=\frac{t}{2}\sum_{l=0}^{N_D} 
\int_0^{2\pi }\frac{\textrm{d}\varphi}{2\pi} \,
\left(P_{n_i,n_i,n_i}(\hat{k})+2 P_{n_i,-n_i,-n_i}(\hat{k})\right)
\\
&\bar{B}_{i,k}= \frac{t}{2}\sum_{l=0}^{N_D}
\int_0^{2\pi }\frac{\textrm{d}\varphi}{2\pi} \,
2\left(P_{n_i,n_k,n_k}(\hat{k})+P_{n_i,-n_k,-n_k}(\hat{k})\right)
\\
&\bar{C}_{i,k}= \frac{t}{2}\sum_{l=0}^{N_D}
\int_0^{2\pi }\frac{\textrm{d}\varphi}{2\pi} \,
P_{-n_k,-n_i,n_k}(\hat{k})
\mbox{,}    
\end{split}
\end{equation*}
where
\begin{equation*}
\begin{split}
& P_{n_1,n_2,n_3}(\hat{k})=
\frac{1}{N_{n_1}^- N_{n_2}^- N_{n_3}^-}+
\frac{1}{N_{n_1}^+ N_{n_2}^+ N_{n_3}^+}
\\
& N^{\pm}_{n}=\omega_l+\imath\left[\pm\mu\bar{B}+\vec{Q}_{n}\hat{k}\right] 
\mbox{.}    
\end{split}
\end{equation*}
In contrast to Eq.~(\ref{eq:g4fin1}) no approximations have 
been used in deriving Eq.~(\ref{eq:fenparfourth}).

Using Eq.~(\ref{eq:condmlatt}) all possible 2D 
lattices and wave vectors may be calculated numerically. 
The stable lattice at $H_{FFLO}(T)$ is then determined 
from the condition of lowest $G_{p}^{(4)}$, taking also the 
LO state into consideration. It turns out, that it is 
energetically favorable at $H_{FFLO}(T)$ if all eigenfunctions 
in the order parameter expansion~(\ref{eq:ffloopans}) have 
equal weight, i.e. $c_i=1$ for all $i$.        

\begin{figure}[htbp]  
\begin{center}\leavevmode
\includegraphics[width=8cm]{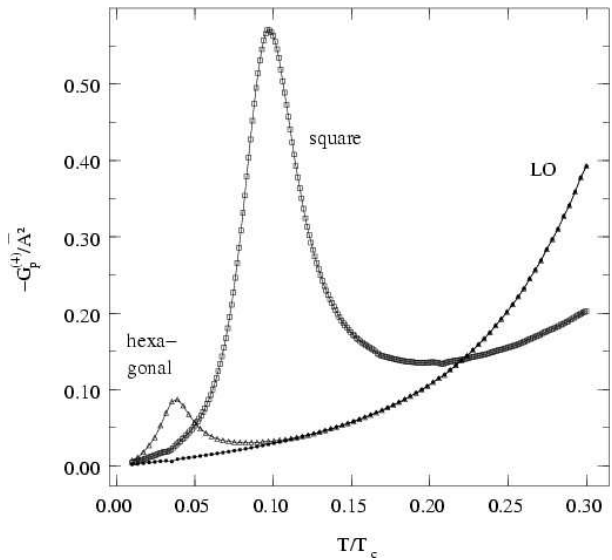}
\caption{\label{fig:sollatbcr} (Reduced image 
quality due to arXiv restrictions) The fourth 
order term $G_{p}^{(4)}$ (minimized with respect to $|\Delta|$) 
divided by $-\bar{A}^2$ [see Eq.~(\ref{eq:fenparsec})] 
at $H_{FFLO}(T)$ for three different periodic structures 
as a function of reduced temperature $t=T/T_c$. The part of 
the hexagonal curve which is lower than the LO state is 
not visible, since the coefficients $c_i$ are determined 
automatically to yield the highest possible solution for
$-G_{p}^{(4)}/\bar{A}^2$. } 
\end{center} 
\end{figure}

The result of the numerical search for the lowest free
energy of periodic structures, characterized by maximal 
three pairs of reciprocal wave vectors, is displayed in 
Fig~\ref{fig:sollatbcr}. The highest curve at a given 
temperature corresponds to the stable lattice. For 
$0.22<t<0.56$ the one-dimensional LO state is realized.
For $t<0.22$ 2D periodic structures appear,
namely the square state for $0.05<t<0.22$, and the 
hexagonal state for $t<0.05$ (we use here the notation 
of Shimahara~\cite{HIR4} for the 2D states).
Besides the fact that the triangular state~\cite{HIR4}
is absent, because it is neither even nor odd, the present 
results agree quantitatively with those of 
Shimahara~\cite{HIR4}, obtained within a different, but 
equivalent, formalism. Thus, more complicated 
2D periodic structures than those found already 
in Ref~\cite{HIR4} do not exist in the considered range of 
temperatures; the assumption of equal weight for different 
wave vectors [$c_i=1$ for all $i$ in Eq.~(\ref{eq:ffloocoeff})] 
has also been confirmed for these states. 

The temperature region \emph{below} $t=0.01$ has been 
investigated recently by Mora and Combescot\cite{MORCOM}. 
They found a series of states characterized by an 
even (total) number $2N=8,10,\ldots$ of different wave vectors, 
all entering the order parameter expansion with equal 
weight, and with $N$ increasing with decreasing 
temperature. Merging these results with the present ones, 
one obtains a very simple description of all of the 
FFLO states at the phase boundary, namely an 
infinite number of states, each one being a linear 
combination of $N=1,2,\ldots$ cosine functions of equal weight 
and with $N$ different, but equally spaced, wave vectors.

Of course, it is also of interest to investigate the 
possible equilibrium structures in the region \emph{below} 
the critical field. As a first step in this direction,
preliminary calculations at $0.95H_{FFLO}$ and $0.90H_{FFLO}$ have been 
performed, using the fourth order 
expansion~(\ref{eq:fenparlim}), which is not valid near 
first order transition lines. The result is surprising and 
shows a revival of the LO state in the low temperature region.

\begin{figure}[htbp]  
\begin{center}\leavevmode
\includegraphics[width=8cm]{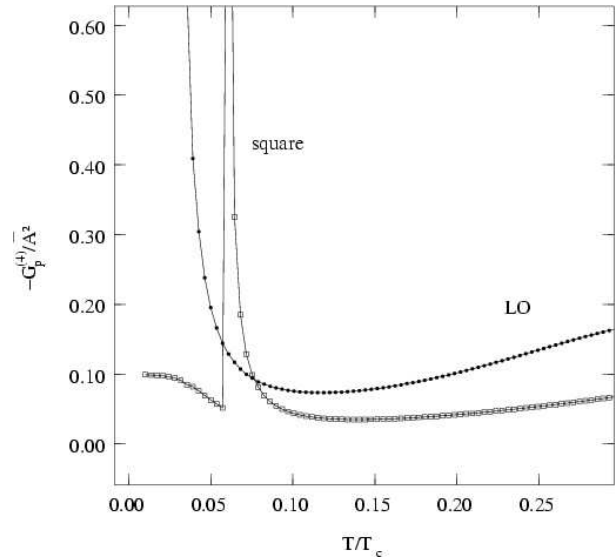}
\caption{\label{fig:sllaa090bcr} (Reduced image 
quality due to arXiv restrictions)  The 
term $-G_{p}^{(4)}/\bar{A}^2$ for the LO state and the square state 
at $0.90H_{FFLO}(T)$, as a function of reduced temperature $T/T_c$. 
The hexagonal curve is lower than the LO state and is 
not displayed in this figure.} 
\end{center} 
\end{figure}

In Fig.~\ref{fig:sllaa090bcr} the terms 
$-G_{p}^{(4)}/\bar{A}^2$, for the three states displayed in 
Fig~\ref{fig:sollatbcr}, are plotted as a function of 
temperature \emph{below} the transition line, at $0.9H_{FFLO}$. 
The hexagonal state (not visible) does not exist any more. 
The usual square state (characterized by $c_i=1$) is only stable 
in a very small temperature interval $0.061<t<0.075$. The 
LO state is now stable in a much larger interval 
$0.075<t<0.56$, as compared to Fig~\ref{fig:sollatbcr}. 
It is also stable in a small temperature region below 
$t=0.061$. But at $t\approx0.017$ the factor $-G_{p}^{(4)}/\bar{A}^2$ 
for the LO state has a singularity and jumps from $+\infty$ 
to $-\infty$. This implies that the fourth order term (for 
the LO state) changes sign and that a first order 
transition occurs somewhere in the vicinity of this 
singularity; higher order terms in the free energy 
would be required for a quantitative treatment. 
Between this singularity at $t\approx0.017$ and the lowest 
considered temperature $t=0.01$ the stable state is 
again characterized by a square unit cell. However, 
the order parameter in this temperature range, $0.01<t<0.017$, 
is given by a linear combination of plane wave states 
[see Eqs.~(\ref{eq:ffloopans}),(\ref{eq:ffloocoeff})] with 
a real coefficient $c_1=1$ and an \emph{imaginary} coefficient 
$c_2=\imath$. The usual order parameter structure for the square 
lattice, which is characterized by two real weight factors 
of equal magnitude ($c_1=c_2=1$), is not equivalent to this 
case and has higher free energy. 

The results below $H_{FFLO}(T)$ indicate, 
that the 2D states are only stable in a tiny 
interval near the phase boundary, and that the 
one-dimensional LO state reappears inside the 
superconducting state. The square state - the one with
the smallest $N$ ($N=2$) - has the largest stability 
region, as one would also expect from the free energy 
balance shown in Fig~\ref{fig:sollatbcr}. We shall come 
back to the  question of the stability of the 2D states 
in section~\ref{sec:tttppr}, considering it from a 
different point of view. The structure found below 
the singular point of the LO state 
(see Fig.~\ref{fig:sllaa090bcr}) raises the question, 
if still other order parameter structures, different from 
those found at the transition line, will appear near $t=0$ 
deep in the superconducting state. The present fourth order
expansion is not really appropriate to answer this question.

\section{\label{sec:tttppr} Transition to the purely 
paramagnetic regime}

The limit $n\to\infty$ of the series of paramagnetic vortex states,
discussed in section~\ref{sec:rffpf}, is now well known; for 
$0.22<t<0.56$ the one-dimensional LO state is realized, while
2D states of square or hexagonal type, predicted  
by Shimahara\cite{HIR4}, appear at lower $t$. The region 
of still smaller $t$, below $t=0.01$, which has been studied 
by Mora and Combescot\cite{MORCOM}, will not be considered 
here. The way, this limit is approached, is, however, unknown. 
Thus, we address ourselves in this section to the the question 
of \emph{how} the one- or two-dimensional unit cell of the 
FFLO state develops from the unit cell of the paramagnetic 
vortex states if the Landau level index $n$ tends to infinity.   

This limiting process is very interesting, because a 
vast number of different states with different symmetry is 
passed through in a small interval of tilt angles $\theta$. 
The unit cell of the \emph{finite}-$n$ states is subject 
to the condition that it carries exactly a single quantum 
of flux of the perpendicular field $B_{\perp}$. Since $B_{\perp}\to0$ as 
$n\to\infty$, at least one of the unit cell vectors must approach 
infinite length - i.e. the dimension of the macroscopic 
sample - in this limit. Thus, the $n\to\infty$ limiting 
process describes a transition from a microscopic (or 
mesoscopic) length scale to a macroscopic length scale.
   
The transition to the FFLO state has previously been 
investigated by Shimahara and Rainer \cite{SHIMARAIN} in 
the linear regime. They found the important relation  
\begin{equation}
  \label{eq:srrelln}
q=\lim_{n \to \infty}\sqrt{4eB_{\perp}n/ \hbar c }
\mbox{,}
\end{equation}
where $q$ is the absolute value of the FFLO wave vector
(here we changed to ordinary units). Eq.~(\ref{eq:srrelln}) 
has been derived by identifying the asymptotic form of the 
Hermite polynomials\cite{GRADRYSH} with the form 
of the LO order parameter. It implies that a relation 
\begin{equation}
  \label{eq:asbobs}
B_{\perp} \approx \frac{\beta}{n},\qquad \beta=\frac{\hbar cq^2}{4e}  
\mbox{}
\end{equation}
holds at large $n$. The validity of 
Eq.~(\ref{eq:srrelln}) may also be checked numerically by 
comparing the numbers $\beta$ and $q$, which are both obtained 
from the upper critical field equation.

Relation~(\ref{eq:asbobs}) may be derived from basic 
physical properties of the present system. The energy 
spectrum for planar Cooper pairs in a perpendicular 
magnetic field $B_\perp$ is the same as for electrons and 
is given by   
\begin{equation}
  \label{eq:ensppbp}
E_n=\hbar\omega(n+\frac{1}{2}),\qquad \omega=\frac{eB_{\perp}}{mc} 
\mbox{.}
\end{equation}
Considering now the energy spectrum of Cooper pairs 
for $B_\perp=0$, one has to distinguish two cases. First, in 
the common situation without a large spin-pair-breaking field, 
all Cooper pairs occupy the lowest possible energy $E=0$, 
which is the kinetic energy $p^2/4m$ taken at the Cooper pair 
momentum $p=0$. Second, if a large spin-pair-breaking 
field parallel to the conducting plane exists, the energy 
value to be occupied by the Cooper pairs, shifts to a 
finite value $p^2/4m$, since the Cooper pairs acquire a 
finite momentum $p$ due to the Fermi level shift discussed
in section~\ref{sec:intro}. Thus, in the latter case, 
which is of interest here, the Landau 
levels~(\ref{eq:ensppbp}) must obey the condition
\begin{equation}
  \label{eq:fcfuopv}
E_n= \frac{\hbar e}{mc}B_{\perp}(n+\frac{1}{2})
\xrightarrow[B_\perp\to 0]{}
\frac{p^2}{4m}   
\mbox{}
\end{equation} 
for $B_\perp\to 0$. If $p$ is replaced by the wave number $q=p/\hbar $, 
Eq.~(\ref{eq:asbobs}) becomes equivalent to 
Eq.~(\ref{eq:fcfuopv}). The limiting behavior expressed by 
Eq.~(\ref{eq:srrelln}) or Eq.~(\ref{eq:asbobs}) is therefore 
a direct consequence of Landau's result for the 
energy eigenvalues of a charged particle in a magnetic 
field. 

Combining Eq.~(\ref{eq:asbobs}) with analytical results at
$T=0$, the limiting behavior of the unit cell as $n \to \infty$ 
may be understood. Expressing the FFLO wave number
in terms of the BCS coherence length $\xi_{0}$ by means of the 
relation $q=(2/\pi)\xi_{0}^{-1}$, and using the flux quantization 
condition in the form 
\begin{equation}
  \label{eq:fqucof}
F^{(n)}B^{(n)}_{\perp}=\Phi_0
\mbox{,}
\end{equation}
[with $\Phi_0=hc/2e$ and $B_{\perp }$ defined by Eq.~(\ref{eq:asbobs})] 
the area $F^{(n)}$ of the unit cell for pairing in Landau 
level $n$ is approximately given by
\begin{equation}
  \label{eq:ainlap}
F^{(n)}  \doteq\pi^3 \xi_0^2 n
\mbox{.}
\end{equation}
Thus, the unit cell area diverges with the first power 
of $n$. The behavior of the magnetic length $L$ , which 
is defined by the relation $B_{\perp}=(\Phi_0/\pi)L^{-2}$, is given by
$L\doteq\pi \xi_{0} n^{1/2}$.    

Eq.~(\ref{eq:ainlap}) is not sufficient to determine the  
shape of the unit cell in the limit of large $n$. 
However, a simple possibility to produce a one-dimensional 
periodic LO structure for $n \to \infty$ is a divergence 
of one of the unit cell lengths, say $b$, of 
the form $b \approx n$, while the second length $a$ remains 
constant, i.e. $a \approx n^{0}$. The numerical results for 
the states referred to in section~\ref{sec:rffpf} 
as ''FFLO-like``, or quasi-one-dimensional states 
show a behavior 
\begin{equation}
  \label{eq:aboaln}
\frac{a}{L}\doteq\frac{\chi}{\sqrt{n}} 
\mbox{,}
\end{equation}
which is in agreement with this possibility. The 
numerical value of the constant $\chi$ is close to $2\sqrt{2}$, 
which corresponds to $a=2\pi\xi_{0}$ and to the lattice constant $\pi /q$ 
of the LO state. Thus, the LO state may be identified als the
limiting case of the quasi-one-dimensional states of
section~\ref{sec:rffpf} for large $n$; the distance of 
the FFLO-lines is essentially independent of $n$, while 
the periodicity length $b\sin\alpha$ in the direction 
perpendicular to the lines tends to infinity 
(like $b\sin\alpha \doteq n\pi/q$) for $n\to\infty$. The one-dimensional FFLO
unit cell is a \emph{substructure} that develops inside 
the diverging unit cell of the paramagnetic vortex 
states.

To complete the description of the transition to the 
LO state, the above lattice structure may be used in 
Eq.~(\ref{eq:opdimlos}) to perform the limit $n \to \infty$ of 
the order parameter expansion $\Delta_{n}$. We consider a 
2D sample of \emph{finite} area $F_p$, which 
contains $N_aN_b$ ''small'' unit cells of size $F_c=ab\sin\alpha$.
The total area is given by $F_p=L_aL_b\sin\alpha$ with $L_a=N_aa$ and 
$L_b=N_bb$. For different $n$ the size and shape of $F_c$ may
change while $F_p$ remains, of course, unchanged. Adopting 
the above model for the behavior of the unit cell as a 
function of $n$, we have $n$-independent numbers $N_a$ 
and $a$, while $b=b^{(n)}$ increases linearly with $n$ and 
$N_b=N_{b}^{(n)}$ decreases consequently according to          
\begin{equation}
  \label{eq:bonibd}
N_{b}^{(n)}=\frac{L_b}{b^{(n)}}=\frac{2L_b\sin\alpha}{\pi^2\xi_0}\,\frac{1}{n}
\mbox{.}
\end{equation}
Thus, a largest possible Landau number $n=n_c$ exists, which 
corresponds to $b^{(n_c)}=L_b$ (or $N_{b}^{(n_c)}=1$) and is given by 
\begin{equation}
  \label{eq:lplqniafs}
n_c=\frac{2L_b\sin\alpha}{\pi^2\xi_0}
\mbox{.}
\end{equation}
This cutoff $n_c$ agrees exactly, in the present model, with 
the number of $n=1$ unit cells fitting into a length $L_b$.     
As an additional consequence of the finite area of the 
sample, only  a finite number of terms occur in the sum 
over $m$ in Eq.~(\ref{eq:opdimlos}). This number is 
fixed by the condition that the ''center positions''  
$y_m=mb\sin\alpha=m\pi L^2/a$, lie inside the sample\cite{PRANSTE}. 
This leads to the condition
\begin{equation}
  \label{eq:condfpm}
-\frac{aL_b\sin\alpha}{2L^2\pi} \leq m < \frac{aL_b\sin\alpha}{2L^2\pi}
\mbox{,}
\end{equation}
which is in the limit $n=n_c$ only fulfilled for $m=0$. Using 
the asymptotic expansion\cite{GRADRYSH} of the Laguerre 
polynomial $L_n$, for large and even $n=2j$, and taking into 
account only the term with $m=0$ in the sum of 
Eq.~(\ref{eq:opdimlos}), the order parameter takes the form  
\begin{equation}
  \label{eq:opasfln}
\Delta_{2j} \approx  A C_{2j} D_{j} \cos \left(\frac{\sqrt{8j}}{L}y \right)
\mbox{.}
\end{equation}
The amplitude $A$ [see Eq.~(\ref{eq:normefho})] is, in the 
present system of (ordinary) units, given by  
\begin{equation*}
A=\left(\frac{2B_{\perp}}{\Phi_0L_a^2} \right)^{1/4}
\mbox{.}
\end{equation*}
The coefficient $C_{2j}$ is, in the limit $n \to n_c$, simply 
given by $C_{2j}=(F_p)^{1/2}$ [see Eq.~(\ref{eq:amplclim})], 
and the coefficient $D_j$ takes the form 
\begin{equation*}
D_j=\frac{2^j (j-1)!}{(-1)^j\sqrt{2\pi(2j-1)!}}
\mbox{.}
\end{equation*}
While these factors, $A,\,D_j,\,C_{2j}$ diverge for $n \to \infty$,  
if the sample dimensions approach infinity, all 
singularities cancel if $n$ is replaced by the cutoff $n_c$, 
and one obtains the expected result, $\Delta_{n_c}=\cos qy$, 
for the one-dimensional periodic order parameter structure 
in the purely paramagnetic limit.

The transition to the \emph{two-dimensional} (square and 
hexagonal) periodic states found by Shimahara\cite{HIR4}
is more involved than the transition to the LO state. Let 
us restrict to the square state, which is the simplest of 
all 2D states, and is also most stable from a thermodynamic 
point of view. 

For the square state, which is a linear combination 
of two LO states with orthogonal wave vectors, one 
would expect a divergent behavior of \emph{both} unit 
cell basis vectors of the type $a \approx n^{1/2}$, $a=b \approx n^{1/2}$. 
Consequently, choosing a square unit cell in the (exact) 
order parameter expansion Eq.~(\ref{eq:opmodsqare}), one 
would expect to find a substructure which becomes increasingly 
similar, with increasing $n$, to the structure of 
Shimaharas square state (line-like order parameter zeros, 
in the form of two sets of orthogonal straight lines and 
circles).  Numerical calculations, performed in the 
range $n<40$ are, however, not in agreement with this 
expectation. 

On the other hand, the mathematical limit of the order 
parameter~(\ref{eq:opmodsqare}) yields in fact a 
2D state with the periodicity of the FFLO wave vector and
square symmetry, as shown in appendix~\ref{sec:tslotop} for
a simplified model. The explanation for this apparent 
contradiction is provided by the result 
[relation~(\ref{eq:arcftln}) of appendix~\ref{sec:tslotop}],
that the quantum number $n$ for a square state must obey 
the condition $n=\pi N^2$, where $N$ is an integer. 
This is a general result, which has been derived using 
essentially only the behavior $a\approx n^{1/2}$ for large $n$. The 
latter is a consequence of the flux quantization condition 
and the shape of the unit cell.     

Of course, the relation $n=\pi N^2$ cannot be fulfilled exactly 
for finite numbers $n,\,N$ (for a sample of finite extension)
since $\pi$ is an irrational number. The proper meaning of 
this relation is, that the sequence of states with 
quantum numbers $n=int(\pi N^2),\;N=1,2,\ldots$ represents a sequence 
of approximations (of increased quality) to the square state.
Thus, the square state is the limit of a sequence defined 
on a very \emph{small} subset of the set of integer numbers.

\begin{figure}[htbp]  
\begin{center}\leavevmode
\includegraphics[width=8cm]{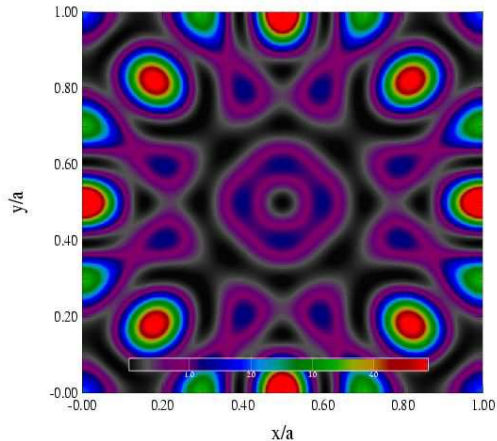}
\caption{\label{fig:cpfni28} (Reduced image 
quality due to arXiv restrictions) Contour 
plot of the square of the order parameter modulus for 
Landau quantum number $n=28$ and a unit cell with 
parameters $a=b,\,\alpha=\pi/2$.} 
\end{center} 
\end{figure}

This explains, why no systematic development of the square
state with increasing $n$ has been observed in the numerical
calculations. The largest quantum number in the considered 
range ($n<40$), which fulfills the above condition is 
$n=28$ (corresponding to $N=3$). The order parameter modulus 
for $n=28$ is shown in Fig.~\ref{fig:cpfni28}. It reveals, 
in fact, a certain similarity to the structure of the 
square state (at least more similarity than any other state 
in the considered range). The arrangement of isolated 
order parameter zeros in Fig.~\ref{fig:cpfni28} 
shows a tendency towards the formation of line-like 
zeros. Clearly, an extremely high $n$ and an extremely 
sharp definition of the tilt angle would be required 
to produce a really good approximation to the square 
state. The final conclusion of the present analysis for 
the square state, that extreme requirements with 
regard to the definition of the tilt angle must be 
fulfilled in order to produce it, will probably hold 
for all other 2D states as well. 

The above analysis of the formation of the FFLO state(s)
as limit(s) of the paramagnetic vortex states for $n \to \infty$ 
has been based on relation~(\ref{eq:fcfuopv}). In addition, 
relation~(\ref{eq:fcfuopv}) allows for an intuitive 
understanding of the unusual phenomenon of Cooper pairing 
at higher $n$, encountered in the present configuration. The 
choice $n=0$ for the ordinary vortex state - in the absence 
of paramagnetic pair-breaking - corresponds to the lowest 
energy the system can achieve for $p=0$. For sufficient 
large $H_{\parallel}$ and decreasing $H_{\perp}$, the Landau level spacing 
becomes smaller than the kinetic energy and the system 
has to perform a quantum jump from the $n=0$ to the 
$n=1$ pairing state, in order to fulfill the requirement 
of given energy as close as possible within the 
available range of discrete states [Inserting $n=1$ 
in Eq.~(\ref{eq:fcfuopv}) determines the angle $\theta_1$ as 
given by Eq.~(\ref{eq:docan1})]. For the same reason, 
a series of successive transitions to superconducting 
states of increasing $n$ takes place with further 
decreasing $H_{\perp}$, until the FFLO state is finally 
reached at $B_{\perp}= 0$. The FFLO state for $n \to \infty$ may 
obviously be considered as the continuum limit, or 
quasiclassical limit, of this series of Cooper 
pair states, which starts with the ordinary vortex 
state at $n=0$.

\section{\label{sec:concln} Conclusion}
The paramagnetic vortex states studied here, appear in 
a small interval of tilt angles close to the 
parallel orientation. A common feature of all of these 
states is a finite momentum of the superconducting 
pair wave function, which is due to the large parallel 
component of the applied magnetic field. In these new 
superconducting states the Cooper pairs occupy quantized 
Landau levels with nonzero quantum numbers $n$. The number 
$n$ increases with decreasing tilt angle and tends to 
infinity for the parallel orientation, where the FFLO 
state is realized. The unusual occupation of higher 
Landau levels may be understood in terms of the finite 
momentum of the Cooper pairs. 

The end points of the infinite series of Cooper-pair
wave states occupying different $n$ are the 
ordinary vortex state at $n=0$ and the FFLO state at 
$n=\infty$. The dominant pair-breaking mechanism in 
the vortex state is the orbital effect, while Cooper 
pairs can only by broken by means of the spin effect 
in the FFLO state. The equilibrium structure of the 
new states, which occupy the levels $0<n<\infty$, is 
very different from the structure of the FFLO state(s), 
despite the fact, that the difference in tilt angles and 
phase boundaries may be small. Generally speaking, the
equilibrium structures of the new states reflect the 
presence of \emph{both} pair-breaking mechanisms; the 
fact that the local magnetic response may be diamagnetic 
or paramagnetic depending on the position in the unit 
cell may be understood in terms of this competition. 
A second unusual property, also closely related to 
the simultaneous presence of both pair-breaking 
mechanisms, is the coexistence of vortices 
and antivortices in a single unit cell.

The FFLO state has been predicted in 1964 and a 
large number of experimental and theoretical works dealing 
with this effect have been published since then. A definite 
experimental verification has not been achieved 
by now. However, recent experiments in the 
organic superconductor $\kappa-(BEDT-TTF)_2Cu(NCS)_2$ and other 
layered materials\cite{NSSBAP,SSNAKD,ZUOBMSW,SHISYASA} 
revealed remarkable agreement\cite{MANKL} with theory, 
both with regard to the angular- and the 
temperature-dependence of the upper critical field. 
In these phase boundary experiments, identification 
of the FFLO precursor states, studied in the present 
paper, seems possible if the tilt angle is defined with 
high precision. To obtain a more direct evidence for 
all of these unconventional states, including the FFLO 
limit, other experiments, such as measurements of the 
local density of states by means of a scanning tunnelling 
microscope would be useful.

%
\begin{acknowledgments}
I would like to thank D.Rainer, Bayreuth and H.Shimahara,
Hiroshima for useful discussions and helpful comments 
during the initial phase of this work. 
\end{acknowledgments}

\appendix

\section{\label{sec:sou} System of units and notation
}
In this appendix we use primes to distinguish  
Eilenbergers dimensionless quantities, which will be 
used in sections~\ref{sec:entucf}-\ref{sec:sotfs}, from 
ordinary ones. The primes will be omitted in 
sections~\ref{sec:entucf}-\ref{sec:sotfs}.
\begin{flushleft}
\textbf{temperature}: $t=T/T_c$ \\ 
\textbf{length}: $\vec{r}\,^{'}=\vec{r}/R_0$,    
$R_0=\hbar v_F/2\pi k_BT_c=0.882\,\xi_0$, $\xi_0$ is the BCS coherence length.\\
\textbf{Fermi velocity}: $\hat{v}_F=\vec{v}_F/v_F$ \\
\textbf{wave number}: $k\,^{'}=kR_0$ \\ 
\textbf{Matsubara frequencies}: $\omega_l^{'}=\omega_l/\pi k_BT_c=(2l+1)t$  \\
\textbf{order parameter}: $\Delta^{'}=\Delta /\pi k_BT_c $  \\
\textbf{magnetic field}: $\vec{H}\,^{'}=\vec{H}/H_0$, where 
$H_0=\hbar c/2eR_0^2$  \\
\textbf{vector potential}: $\vec{A}\,^{'}=\vec{A}/A_0$, where
$A_0=\hbar c/2eR_0$   \\
\textbf{magnetic moment}: $\mu^{'}=\mu/\mu_0=\pi k_BT_c/mv_F^2$, 
where $\mu_0=\pi k_B T_c/H_0$. Note that the dimensionless magnetic 
moment $\mu^{'}$ agrees with the quasiclassical parameter. \\
\textbf{Gibbs free energy}: $G^{'}=G/[\left(\pi k_B T_c 
\right)^2 N_F R_0^3]$   
\end{flushleft}
Eilenbergers parameter $\tilde{\kappa}$ is related to the 
GL-parameter $\kappa_0$ of a clean superconductor according 
to the relation
$\tilde{\kappa}=\left(\frac{7}{18}\zeta(3)\right)^{1/2} \kappa_0=
0.6837\kappa_0$.

The symbol $\hat{k}$ denotes a dimensionless, 
2D unit vector. The Fermi-surface average 
of a $\hat{k}-$dependent quantity $a(\hat{k})$ is denoted 
by $\overline{a}$. For our cylindrical Fermi surface 
this average is simply an integral from $0$ to $2\pi$ over 
the azimuth angle $\varphi$. 
\begin{equation*}
  \label{eq:nolblftf}
\overline{a}=\frac{1}{4\pi}\oint \textrm{d}^2\hat{k}\,a(\hat{k})
=\frac{1}{2\pi}\int_0^{2\pi } \textrm{d}\varphi \, a(\hat{k}(\varphi)) 
\mbox{.} 
\end{equation*}

Finally, the symbol $\langle a\rangle$, defined by 
\begin{equation*}
  \label{}
\langle a\rangle=\frac{1}{F_c}\int_{unit\;cell}\mathrm{d}^2r\,a(\vec{r})
\mbox{,} 
\end{equation*}  
denotes a spatial average of a quantity $a(\vec{r})$ over 
a unit cell of area $F_c$.

\section{\label{sec:gapcorfun} Gap correlation function}

It is convenient to express the gap correlation function, 
defined by Eq.~(\ref{eq:defgapcorfun}) in terms of center 
of mass coordinates $\vec{R}=(\vec{r}_1+\vec{r}_2)/2,\,\vec{r}=
\vec{r}_1-\vec{r}_2$, using the notation 
$V^{CM}(\vec{R},\vec{r})=V(\vec{r}_1,\vec{r}_2)$. The function 
$V^{CM}(\vec{R},\vec{r})$ is invariant under center of mass 
translations $\vec{R}\Rightarrow\vec{R}+l\vec{a}+j\vec{b}$ and 
may consequently be expanded in a Fourier series, using 
reciprocal lattice vectors
$\vec{Q}_{l,j}=l\vec{Q}_1+j\vec{Q}_2,\;l,j=0,\pm1,\pm2,\ldots$, 
with basis vectors 
\begin{equation*}
\vec{Q}_1=\frac{2\pi}{a}   
\left(
\begin{array}{c}
\;\;1\\-\frac{1}{\tan\alpha} 
\end{array}
\right),\;\;
\vec{Q}_2=\frac{2\pi}{b}
\left(
\begin{array}{c}
0\\ \frac{1}{\sin\alpha} 
\end{array}
\right)
\mbox{.}
\end{equation*}
The Fourier coefficients of $V^{CM}(\vec{R},\vec{r})$ are 
denoted by $V_{l,j}(\vec{r})$. The Fourier transform of 
$V_{l,j}(\vec{r})$ with respect to $\vec{r}$ is 
denoted by $V_{l,j}^{(p)}(\vec{p})$.

Using the behavior of the gap $\Delta(\vec{r})$ under lattice
translations $\vec{r}\Rightarrow\vec{r}+\vec{r}_{l,j}$, where 
$\vec{r}_{l,j}=l\vec{a}+j\vec{b}$, the important relation
\begin{equation}
  \label{eq:delrieurel}
V_{-l,j}(\vec{r})=\mathrm{e}^{\imath\pi l (j+ \frac{b}{a} \cos\alpha ) }V_{0,0}(\vec{r}+\vec{r}_{j,l})
\mbox{}
\end{equation}
may be proven. This relation, first 
reported by Delrieu\cite{DELRIEU}, shows that all Fourier 
coefficients are known if $V_{0,0}$ is known. A similar 
relation holds for the Fourier transform $V_{l,j}^{(p)}$:  
\begin{equation*}
  \label{eq:delrieutrans}
V_{l,j}^{(p)}(\vec{p})=\mathrm{e}^{\imath \vec{p}\vec{r}_{j,-l}- 
\imath\pi l (j+ \frac{b}{a} \cos\alpha ) }
V_{0,0}^{(p)}(\vec{p})
\mbox{.}
\end{equation*}
The functions $V_{l,j}$ and 
$V_{l,j}^{(p)}$, which are most useful for the evaluation 
of the free energy, may be calculated by proceeding along 
the chain
\begin{equation*}
  \label{eq:achainfc}
V^{CM}(\vec{R},\vec{r})
\Rightarrow V_{0,0}(\vec{r})
\Rightarrow V_{0,0}^{(p)}(\vec{p})
\Rightarrow V_{l,j}^{(p)}(\vec{p})
\Rightarrow V_{l,j}(\vec{r})
\mbox{,} 
\end{equation*}
where an arrow denotes either calculation of a 
Fourier coefficient, or of a Fourier transform,
or application of Delrieu's relation.  

Using the order parameter expansion~(\ref{eq:opdimlos}) and 
performing the necessary manipulations, the 
result for $V_{l,j}^{(p)}$ is given by
\begin{equation}
\label{eq:frffcft}
\begin{split}
V_{l,j}^{(p)}(\vec{p})=&\frac{4\pi}{\bar{B}_{\perp}}
(-1)^{n+lj}\langle|\Delta_n|^2\rangle
\mathrm{e}^{-\frac{\vec{p}^2}{\bar{B}_{\perp}}}
L_n\left(\frac{2}{\bar{B}_{\perp}}\vec{p}^{\,2} \right)\cdot \\
&\mathrm{e}^{-\imath \pi n \frac{b}{a}\cos\alpha}
\mathrm{e}^{\imath\frac{F_C}{2\pi}
\left(p_xQ_{l,j,y}- p_yQ_{l,j,x}\right)}
\mbox{,}
\end{split}
\end{equation}
where $Q_{l,j,x},\,Q_{l,j,y}$ are the $x$ and $y$ components 
respectively of the reciprocal lattice vector $\vec{Q}_{l,j}$.   
The final result for $V_{l,j}$ is given by 
\begin{align}
&V_{l,j}(\vec{r})= (-1)^{lj}\langle|\Delta_n|^2\rangle \cdot  \nonumber  \\ 
&\mathrm{e}^{-\imath \pi n \frac{b}{a}\cos\alpha}
\mathrm{e}^{-\frac{\bar{B}_{\perp}}{4}G^2_{l,j} }
L_n\left(\frac{\bar{B}_{\perp}}{2}G^2_{l,j} \right)
\mbox{,}  \label{eq:frffspa} \\
&\frac{\bar{B}_{\perp}}{2}G^2_{l,j}=\frac{\pi}{F_C}\vec{r}^{\,2}
+xQ_{l,j,y}-yQ_{l,j,x}+x_{l,j}
\mbox{,} \nonumber \\
&x_{l,j}=\frac{\pi}{\sin\alpha}
\left[\left(\frac{b}{a}\right)l^2+
\left(\frac{a}{b}\right)j^2-2lj\cos\alpha  \right]
\mbox{.} \label{eq:affs6k}
\end{align}

The usefulness of the gap correlation function for 
pair-wave states with with arbitrary $n$ is essentially 
based on the translational invariance of the observable 
quantities $|\psi|$ and $B$.

\section{\label{sec:tgll} The Ginzburg-Landau limit}
Let us first consider the upper critical field $H_{c2}^{GL}$,
which is determined by $\bar{G}^{(2)}=0$, for 
$\mu=0,\,n=0$ ($H_{\parallel}=0$) and $t\to1$. Solving this 
equation in this limit, one finds, using 
ordinary units,     
\begin{equation}
  \label{eq:myglucf}
H_{c2}^{GL}=1.222\frac{\Phi_0}{2\pi\xi_0^2}(1-t)
\mbox{.}
\end{equation}
Eq.~(\ref{eq:myglucf}) differs from the usual GL result 
by a factor of $3/2$. This discrepancy is due to our 
use of a cylindrical Fermi surface, instead of a 
spherical one, and can be eliminated by replacing 
the GL parameter $\kappa$ by $3\kappa_{GL}/2$ (the quantities used 
in Eilenberger units are derived assuming a spherical 
Fermi surface).

The magnetization relation~(\ref{eq:marebh}) takes the 
following form for $H_{\parallel}=0,\,\mu=0,\,t\to1$:
\begin{equation}
\label{eq:amrpwf}
\bar{B}-H=4\pi M=\frac{H-H_{c2}}{2\tilde{\kappa}^2/A_{\perp}-1}
\mbox{.}
\end{equation}
The coefficient $A_{\perp}$ in~(\ref{eq:amrpwf})is given 
by~(\ref{eq:scas3}). The fourth order free energy 
contribution~(\ref{eq:g4fin1}) takes the form
\begin{equation}
  \label{eq:focigll}
\bar{G}^{(4)}=
\frac{S^{(1)}}{4} \left[
 \sum_{l,m} f_1^2(x_{l,m}) -\frac{S^{(1)}}{4\tilde{\kappa}^2} \sum_{l,m}^{\prime} f_1^2(x_{l,m})
 \right]
\mbox{,}
\end{equation}
where $S^{(1)}=7\zeta(3)/8$. The first sum in Eq.~(\ref{eq:focigll})
turns out to agree with Abrikosov's geometrical 
factor $\beta_A$, 
\begin{equation}
\label{eq:gfag3h}
\sum_{l,m} f_1^2(x_{l,m})=\beta_A
\mbox{,}
\end{equation}
as discussed in more detail in KRS\cite{KLRAISHI}.
Performing again the above replacement of $\kappa$
one arrives at Abrikosovs well-known result
\begin{equation}
  \label{eq:abwnr}
4\pi\frac{\partial M}{\partial H}\Big|_{H_{c2}}=\frac{1}{(2\kappa^2_{GL}-1)\beta_A}
\mbox{.}
\end{equation}
Eq.~(\ref{eq:gfag3h}) remains also valid for $n>0$. For 
the nonmagnetic terms in Eq.~(\ref{eq:g4fin1}), the 
Matsubara sum $S_{l,m}^{(1)}$ may be considered as a low 
temperature correction to the GL term~(\ref{eq:gfag3h}).
The GL-limit of the local magnetic field 
$B_{1\perp}$ [see Eq.~(\ref{eq:b1senk})] has 
also been calculated and has been found to obey 
the correct GL relation\cite{SAJATH} between magnetic 
field and square of order parameter. Here, the 
low-temperature corrections are contained in the 
Matsubara sum $S_{l,m}^{(2)}$.

\section{\label{sec:tlotovl} The limit of the ordinary 
                           vortex lattice}
It is of interest to investigate the limit of 
Eq.~(\ref{eq:g4fin1}) corresponding to the ordinary vortex 
lattice. We consider a situation without 
paramagnetic pair-breaking, i.e. set $\mu=0,\, \Theta=\pi/2$, and 
ask for the equilibrium structure of the vortex lattice 
and the critical value of $\kappa$ separating type I from type II
superconductivity. To compare with the usual notation,
we use here the same scaling $\kappa \Rightarrow 2\kappa/3$ of the GL parameter 
as in appendix~\ref{sec:tgll}. Figure~\ref{fig:ordvlatt}
shows the free energy $G^{(4)}$ as a function of $a/L,\,\alpha$ for 
$\kappa=1.46$ ($\tilde{\kappa}=1.5$) at $t=0.5$. The flat 
minimum of $G^{(4)}$ at $a/L=1.905,\,\alpha=60$ indicates that the 
stable configuration is, as expected, a triangular vortex 
lattice. No other local minimum of the free energy exists. 
With decreasing $\kappa$ this minimum changes quickly into 
a maximum; below $\kappa\doteq1.36$ the free energy  has no minimum 
at all which means that no spatially varying 
superconducting state exists. The critical value of 
$\kappa\doteq1.36$ separating type I from type II behavior at 
$t=0.5$ agrees fairly well with the result of $\kappa\doteq1.25$ 
obtained by Kramer\cite{KRAZPH} for the phase boundary 
between typ II and type II/1 behavior. For lower 
temperature the agreement is worse; at $t=0.2$ the 
present theory gives $\kappa=2.5$ while Kramer's 
theory\cite{KRAZPH} gives $\kappa=1.7$. Recall that the 
error induced by the asymptotic approximation of 
subsection~\ref{subsec:fourthordcon} increases with 
decreasing temperature.

\begin{figure}[htbp] 
\begin{center}\leavevmode
\includegraphics[width=8cm]{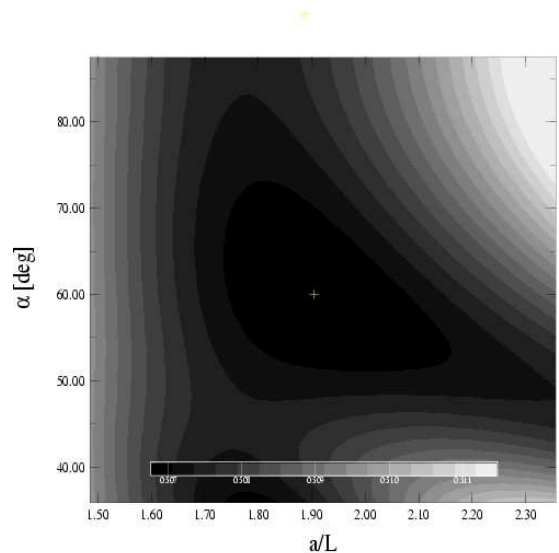}
\caption{\label{fig:ordvlatt} (Reduced image 
quality due to arXiv restrictions)  Contour plot of the 
free energy $G^{(4)}$ as a function of $a/L$ and
$\alpha$ without paramagnetic pair-breaking. Parameters are 
$\tilde{\kappa}=1.5,\;t=0.5,\;\mu=0,\;\Theta=\pi/2$ . The   
minimal value $G^{(4)}=0.5067$ is at $a/L=1.905,\,\alpha=60$.}
\end{center} 
\end{figure}

\section{\label{sec:tslotop} The square limit for a model order
                            parameter}

The square of the the order parameter modulus,
Eq.~(\ref{eq:opmodsqare}), for a square lattice may be 
written in the form
\begin{align}
& |\psi_{n}|^2(x,y,a)=  \sum_{l,j} H_{l,j} 
\mbox{,}
\label{eq:opmodsqst} \\
& H_{l,j}= (-1)^{lj} \textrm{e}^{-\frac{\pi}{2}(l^{2}+j^{2})}
L_n(\pi(l^{2}+j^{2}))\textrm{e}^{\imath\frac{2\pi}{a}(lx+jy)} \nonumber
\mbox{.}
\end{align}
We are interested in the limiting behavior of 
Eq.~(\ref{eq:opmodsqst}) for $n \to \infty $, $a\approx n^{\frac{1}{2}} \to \infty$.
In this limit, the quantity $2\pi/a$ tends to zero and 
the double sum may be approximated by a double integral.
An appropriate tool to  perform such a calculation for 
infinite sums in a systematic way is Poissons summation 
formula. Using a two-dimensional version, which is 
derived in exactly the same way as for single sums, 
Eq.~(\ref{eq:opmodsqst}) may be written in the form
\begin{equation}
\begin{split}
  \label{eq:psffop}
&|\psi_{n}|^2(x,y,a)=\left(\frac{a}{2\pi} \right)^2
\sum_{m_{x}}\sum_{m_{y}}\int\mathrm{d}k^{\prime}_{x}\,\int\mathrm{d}k^{\prime}_{y}\cdot \\
&\textrm{e}^{-\imath\left(m_xk^{\prime}_{x}+m_yk^{\prime}_{y}\right)}
h(k^{\prime}_{x},x,k^{\prime}_{y},y,a)
\mbox{,}
\end{split}
\end{equation}
where $h(k_{x},x,k_{y},y,a)$ is a function representing 
$H_{l,j}$. The problem here is the factor $(-1)^{lj}$
[see Eq.~(\ref{eq:opmodsqst})], which    
must be represented by an infinite series 
of step functions\cite{WALKER}. 

Since we are more interested in the question \emph{if}
a limit with the correct periodicity and symmetry exists, 
than in the detailed functional form of this limit, we 
represent the factor $(-1)^{lj}$ approximately by the real 
part of $\textrm{exp}\imath a^2k_xk_y/4\pi$, i.e. we use the function   
\begin{equation}
  \label{eq:defunhappr}
h(k_{x},x,k_{y},y,a)=H_{l,j}\Big|_{l=\frac{a}{2\pi}k_x,\,j=\frac{a}{2\pi}k_y}
\end{equation}
to represent $H_{l,j}$. Using this model, the absolute 
value of the r.h.s. of Eq.~(\ref{eq:psffop}) will be 
denoted by $S(x,y,a)$ instead of $|\psi_{n}|^2(x,y,a)$. It 
takes the form 
\begin{equation}
\begin{split}
  \label{eq:adfavorhs}
&S(x,y,a)=\left(\frac{a}{2\pi} \right)^2  \Big\arrowvert
\sum_{m_{x}}\sum_{m_{y}}\int\mathrm{d}k^{\prime}_{x}\,\int\mathrm{d}k^{\prime}_{y}\cdot \\
&h(k^{\prime}_{x},x-m_xa,k^{\prime}_{y},y-m_ya,a) \Big\arrowvert
\mbox{,}
\end{split}
\end{equation}
where $h(k_{x},x,k_{y},y,a)$ is given by 
\begin{equation*}
\begin{split}
& h(k_{x},x,k_{y},y,a)=\cos\left(\frac{a^2}{4\pi}k_xk_y \right)
\textrm{e}^{-\frac{a^2}{8\pi}\left(k_y^2+k_y^2\right)}\cdot
\\
&L_n\left(\frac{a^2}{4\pi}\left(k_y^2+k_y^2\right) \right)
\textrm{e}^{\imath(k_xx+k_yy)} 
\mbox{.}    
\end{split}
\end{equation*}
In order to perform the integrations, the relation
\begin{equation*}
  \label{eq:afftfotlp}
L_n(x+y)=\frac{1}{(-1)^n 2^{2n}n!}\sum_{m=0}^{n}{n \choose m}
H_{2m}(\sqrt{x})H_{2n-2m}(\sqrt{y})
\mbox{}
\end{equation*}
may be used to rewrite the Laguerre polynomial in 
the integrand as a sum of products depending on $k_x$ and $k_y$ 
separately. Then, the  integration over $k_x$ may be 
performed\cite{GRADRYSH} and, after a simple shift of the 
integration variable, a second relation\cite{BUCHHOLTZ}   
\begin{equation*}
  \label{eq:asrftbp}
\begin{split}
&(-2)^n H_n(\frac{x+y}{\sqrt{2}})H_n(\frac{x-y}{\sqrt{2}})=\\
& \sum_{m=0}^{n} (-1)^m {n \choose m}
H_{2m}(\sqrt{x})H_{2n-2m}(\sqrt{y})
\mbox{}
\end{split}
\end{equation*}
may be used to calculate the sum over $m$. Performing the 
integration over $k_x$ one obtains the final result
\begin{equation}
\begin{split}
  \label{eq:dfrfdsm}
&S(x,y,a)=\Big\arrowvert \frac{\imath^n}{\sqrt{2}2^nn!} 
\sum_{m_{x}}\sum_{m_{y}} \textrm{e}^{-\frac{\pi}{a^2}\tilde{x}^2}
H_n\left(\frac{\sqrt{2\pi}}{a}\tilde{x}\right) 
\textrm{e}^{-\frac{\pi}{a^2}\tilde{y}^2} \\
&H_n\left(\frac{\sqrt{2\pi}}{a}\tilde{y}\right)  
\left(\textrm{e}^{-\imath \frac{2\pi}{a^2}\tilde{x}\tilde{y}}+
(-1)^n \textrm{e}^{\imath \frac{2\pi}{a^2}\tilde{x}\tilde{y}}   \right) \Big\arrowvert
\mbox{,}
\end{split}
\end{equation}
where the abbreviations $\tilde{x}=x-m_xa,\,\tilde{y}=y-m_ya$ 
have been used. 

We are interested in the limiting value of Eq.~(\ref{eq:dfrfdsm})
for $n \to \infty$. The asymptotic behavior of the Hermite
polynomials\cite{GRADRYSH} for large $n$ implies  
\begin{equation}
\begin{split}
  \label{eq:abohpfln}
H_n\left(\frac{\sqrt{2\pi}}{a}\tilde{x}\right)\approx&
cos\Big[(2n+1)^{1/2}\frac{\sqrt{2\pi}}{a}x-\\
& (2n+1)^{1/2}\sqrt{2\pi}m_x\Big]
\mbox{.}
\end{split}
\end{equation}
Since $a\approx n^{1/2}$, the factor in front of $x$ in 
Eq.~(\ref{eq:abohpfln}) remains finite for $n \to \infty $ and 
defines the FFLO wave vector $q$, i.e.    
\begin{equation}
  \label{eq:tdothfwv}
(2n+1)^{1/2}\frac{\sqrt{2\pi}}{a}=q,\;\mbox{for }n\to\infty 
\mbox{.}
\end{equation}
Eq.~(\ref{eq:tdothfwv}) implies a 
restriction on the possible quantum numbers $n$ of a 
square FFLO state. The fact that an integer number $N$ of 
wave lengths $\lambda=2\pi/q$ must fit into a length $a$, implies 
the condition 
\begin{equation}
  \label{eq:arcftln}
n=\pi N^2
\mbox{}
\end{equation}
in the limit $a \to \infty$. Condition~(\ref{eq:arcftln}) is of 
a general nature and not a specific feature of our model.
For quantum numbers $n$ obeying Eq.~(\ref{eq:arcftln}),
all $m_x,m_y-$dependent phase factors in Eq.~(\ref{eq:abohpfln})
become multiples of $2\pi$ and may be omitted. The limit 
of $S(x,y,a)$ for $a \to \infty$ obtained in this way is well-defined,
i.e. independent of any cutoff, and is given by
\begin{equation}
  \label{eq:titfrfs}
\lim_{n \to \infty}S(x,y,a) \approx  |\cos(qx)\cos(qy)|
\mbox{.}
\end{equation}
A rotation of $\pi/4$ transforms Eq.~(\ref{eq:titfrfs}) into 
the more familiar form\cite{HIR4} $|\cos(q^{\prime}x^{\prime})+cos(q^{\prime}y^{\prime})|$. 
The expected correct result for $|\psi_{n}|^2$ is the 
\emph{square} of the r.h.s. of Eq.~(\ref{eq:titfrfs}). Thus, 
the result of our model calculation differs from the exact 
result. A limiting state of the correct periodicity and 
symmetry has, however, been obtained. 


\bibliography{fflobig.bib}

\begin{thebibliography}{48}
\expandafter\ifx\csname natexlab\endcsname\relax\def\natexlab#1{#1}\fi
\expandafter\ifx\csname bibnamefont\endcsname\relax
  \def\bibnamefont#1{#1}\fi
\expandafter\ifx\csname bibfnamefont\endcsname\relax
  \def\bibfnamefont#1{#1}\fi
\expandafter\ifx\csname citenamefont\endcsname\relax
  \def\citenamefont#1{#1}\fi
\expandafter\ifx\csname url\endcsname\relax
  \def\url#1{\texttt{#1}}\fi
\expandafter\ifx\csname urlprefix\endcsname\relax\def\urlprefix{URL }\fi
\providecommand{\bibinfo}[2]{#2}
\providecommand{\eprint}[2][]{\url{#2}}

\bibitem[{\citenamefont{Fulde and Ferrell}(1964)}]{FUFE}
\bibinfo{author}{\bibfnamefont{P.}~\bibnamefont{Fulde}} \bibnamefont{and}
  \bibinfo{author}{\bibfnamefont{R.~A.} \bibnamefont{Ferrell}},
  \bibinfo{journal}{Phys.\ Rev.} \textbf{\bibinfo{volume}{135}},
  \bibinfo{pages}{A550} (\bibinfo{year}{1964}).

\bibitem[{\citenamefont{Larkin and Ovchinnikov}(1965)}]{LAOV}
\bibinfo{author}{\bibfnamefont{A.~I.} \bibnamefont{Larkin}} \bibnamefont{and}
  \bibinfo{author}{\bibfnamefont{Y.~N.} \bibnamefont{Ovchinnikov}},
  \bibinfo{journal}{Sov.\ Phys.\ JETP.} \textbf{\bibinfo{volume}{20}},
  \bibinfo{pages}{762} (\bibinfo{year}{1965}).

\bibitem[{\citenamefont{Bulaevskii}(1974)}]{BULATILT}
\bibinfo{author}{\bibfnamefont{L.~N.} \bibnamefont{Bulaevskii}},
  \bibinfo{journal}{Sov.\ Phys.\ JETP} \textbf{\bibinfo{volume}{38}},
  \bibinfo{pages}{634} (\bibinfo{year}{1974}).

\bibitem[{\citenamefont{Shimahara and Rainer}(1997)}]{SHIMARAIN}
\bibinfo{author}{\bibfnamefont{H.}~\bibnamefont{Shimahara}} \bibnamefont{and}
  \bibinfo{author}{\bibfnamefont{D.}~\bibnamefont{Rainer}},
  \bibinfo{journal}{J.\ Phys.\ Soc.\ Jpn.} \textbf{\bibinfo{volume}{66}},
  \bibinfo{pages}{3591} (\bibinfo{year}{1997}).

\bibitem[{\citenamefont{Chandrasekhar}(1962)}]{CHANDRA}
\bibinfo{author}{\bibfnamefont{B.~S.} \bibnamefont{Chandrasekhar}},
  \bibinfo{journal}{Appl.\ Phys.\ Lett.} \textbf{\bibinfo{volume}{1}},
  \bibinfo{pages}{7} (\bibinfo{year}{1962}).

\bibitem[{\citenamefont{Clogston}(1962)}]{CLOGSTON}
\bibinfo{author}{\bibfnamefont{A.~M.} \bibnamefont{Clogston}},
  \bibinfo{journal}{Phys.\ Rev.\ Lett.} \textbf{\bibinfo{volume}{9}},
  \bibinfo{pages}{266} (\bibinfo{year}{1962}).

\bibitem[{\citenamefont{Burkhardt and Rainer}(1994)}]{BURKRAIN}
\bibinfo{author}{\bibfnamefont{H.}~\bibnamefont{Burkhardt}} \bibnamefont{and}
  \bibinfo{author}{\bibfnamefont{D.}~\bibnamefont{Rainer}},
  \bibinfo{journal}{Ann.\ Physik} \textbf{\bibinfo{volume}{3}},
  \bibinfo{pages}{181} (\bibinfo{year}{1994}).

\bibitem[{\citenamefont{Klein et~al.}(2000)\citenamefont{Klein, Rainer, and
  Shimahara}}]{KLRAISHI}
\bibinfo{author}{\bibfnamefont{U.}~\bibnamefont{Klein}},
  \bibinfo{author}{\bibfnamefont{D.}~\bibnamefont{Rainer}}, \bibnamefont{and}
  \bibinfo{author}{\bibfnamefont{H.}~\bibnamefont{Shimahara}},
  \bibinfo{journal}{J.\ Low\ Temp.\ Phys.} \textbf{\bibinfo{volume}{118}},
  \bibinfo{pages}{91} (\bibinfo{year}{2000}).

\bibitem[{\citenamefont{M.Rasolt and Tesanovic}(1992)}]{RATEREV}
\bibinfo{author}{\bibnamefont{M.Rasolt}} \bibnamefont{and}
  \bibinfo{author}{\bibfnamefont{Z.}~\bibnamefont{Tesanovic}},
  \bibinfo{journal}{Rev.\ Mod.\ Phys.} \textbf{\bibinfo{volume}{64}},
  \bibinfo{pages}{709} (\bibinfo{year}{1992}).

\bibitem[{\citenamefont{Ryan and Rajagopal}(1993)}]{RYARAJ1}
\bibinfo{author}{\bibfnamefont{J.~C.} \bibnamefont{Ryan}} \bibnamefont{and}
  \bibinfo{author}{\bibfnamefont{A.~K.} \bibnamefont{Rajagopal}},
  \bibinfo{journal}{Phys.\ Rev.} \textbf{\bibinfo{volume}{B47}},
  \bibinfo{pages}{8843} (\bibinfo{year}{1993}).

\bibitem[{\citenamefont{Akera et~al.}(1991)\citenamefont{Akera, MacDonald,
  Girvin, and Norman}}]{AKDOGINO}
\bibinfo{author}{\bibfnamefont{H.}~\bibnamefont{Akera}},
  \bibinfo{author}{\bibfnamefont{A.~H.} \bibnamefont{MacDonald}},
  \bibinfo{author}{\bibfnamefont{S.~M.} \bibnamefont{Girvin}},
  \bibnamefont{and} \bibinfo{author}{\bibfnamefont{M.~R.}
  \bibnamefont{Norman}}, \bibinfo{journal}{Phys.\ Rev.\ Lett.}
  \textbf{\bibinfo{volume}{67}}, \bibinfo{pages}{2375} (\bibinfo{year}{1991}).

\bibitem[{\citenamefont{Tesanovic et~al.}(1989)\citenamefont{Tesanovic,
  M.Rasolt, and L.Xing}}]{TERAXI}
\bibinfo{author}{\bibfnamefont{Z.}~\bibnamefont{Tesanovic}},
  \bibinfo{author}{\bibnamefont{M.Rasolt}}, \bibnamefont{and}
  \bibinfo{author}{\bibnamefont{L.Xing}}, \bibinfo{journal}{Phys.\ Rev.\ Lett.}
  \textbf{\bibinfo{volume}{63}}, \bibinfo{pages}{2425} (\bibinfo{year}{1989}).

\bibitem[{\citenamefont{Rieck et~al.}(1990)\citenamefont{Rieck, Scharnberg, and
  Klemm}}]{RIESCHKLE}
\bibinfo{author}{\bibfnamefont{C.~T.} \bibnamefont{Rieck}},
  \bibinfo{author}{\bibfnamefont{K.}~\bibnamefont{Scharnberg}},
  \bibnamefont{and} \bibinfo{author}{\bibfnamefont{R.~A.} \bibnamefont{Klemm}},
  \bibinfo{journal}{Physica C} \textbf{\bibinfo{volume}{170}},
  \bibinfo{pages}{195} (\bibinfo{year}{1990}).

\bibitem[{\citenamefont{Nicopoulos and Kumar}(1991)}]{NICKUM}
\bibinfo{author}{\bibfnamefont{V.~N.} \bibnamefont{Nicopoulos}}
  \bibnamefont{and} \bibinfo{author}{\bibfnamefont{P.}~\bibnamefont{Kumar}},
  \bibinfo{journal}{Phys.\ Rev.} \textbf{\bibinfo{volume}{B44}},
  \bibinfo{pages}{12080} (\bibinfo{year}{1991}).

\bibitem[{\citenamefont{Manalo and Klein}(2002)}]{MANKL2}
\bibinfo{author}{\bibfnamefont{S.}~\bibnamefont{Manalo}} \bibnamefont{and}
  \bibinfo{author}{\bibfnamefont{U.}~\bibnamefont{Klein}},
  \bibinfo{journal}{Phys.\ Rev.} \textbf{\bibinfo{volume}{B65}},
  \bibinfo{pages}{144510} (\bibinfo{year}{2002}).

\bibitem[{\citenamefont{Gloos et~al.}(1993)\citenamefont{Gloos, Modler,
  Schimanski, Bredl, Geibel, Steglich, Buzdin, Sato, and
  Komatsubara}}]{GMSBGSBSK}
\bibinfo{author}{\bibfnamefont{K.}~\bibnamefont{Gloos}},
  \bibinfo{author}{\bibfnamefont{R.}~\bibnamefont{Modler}},
  \bibinfo{author}{\bibfnamefont{H.}~\bibnamefont{Schimanski}},
  \bibinfo{author}{\bibfnamefont{C.~D.} \bibnamefont{Bredl}},
  \bibinfo{author}{\bibfnamefont{C.}~\bibnamefont{Geibel}},
  \bibinfo{author}{\bibfnamefont{F.}~\bibnamefont{Steglich}},
  \bibinfo{author}{\bibfnamefont{A.~I.} \bibnamefont{Buzdin}},
  \bibinfo{author}{\bibfnamefont{N.}~\bibnamefont{Sato}}, \bibnamefont{and}
  \bibinfo{author}{\bibfnamefont{T.}~\bibnamefont{Komatsubara}},
  \bibinfo{journal}{Phys.\ Rev.\ Lett.} \textbf{\bibinfo{volume}{70}},
  \bibinfo{pages}{501} (\bibinfo{year}{1993}).

\bibitem[{\citenamefont{Eilenberger}(1968)}]{EILE1}
\bibinfo{author}{\bibfnamefont{G.}~\bibnamefont{Eilenberger}},
  \bibinfo{journal}{Z.\ Physik} \textbf{\bibinfo{volume}{214}},
  \bibinfo{pages}{195} (\bibinfo{year}{1968}).

\bibitem[{\citenamefont{Larkin and Ovchinnikov}(1969)}]{LAROVC}
\bibinfo{author}{\bibfnamefont{A.~I.} \bibnamefont{Larkin}} \bibnamefont{and}
  \bibinfo{author}{\bibfnamefont{Y.~N.} \bibnamefont{Ovchinnikov}},
  \bibinfo{journal}{Sov.\ Phys.\ JETP.} \textbf{\bibinfo{volume}{28}},
  \bibinfo{pages}{1200} (\bibinfo{year}{1969}).

\bibitem[{\citenamefont{Alexander et~al.}(1985)\citenamefont{Alexander,
  Orlando, Rainer, and Tedrow}}]{AORT}
\bibinfo{author}{\bibfnamefont{J.~A.~X.} \bibnamefont{Alexander}},
  \bibinfo{author}{\bibfnamefont{T.~P.} \bibnamefont{Orlando}},
  \bibinfo{author}{\bibfnamefont{D.}~\bibnamefont{Rainer}}, \bibnamefont{and}
  \bibinfo{author}{\bibfnamefont{P.~M.} \bibnamefont{Tedrow}},
  \bibinfo{journal}{Phys.\ Rev.} \textbf{\bibinfo{volume}{B31}},
  \bibinfo{pages}{5811} (\bibinfo{year}{1985}).

\bibitem[{\citenamefont{Sarma}(1963)}]{SARMA}
\bibinfo{author}{\bibfnamefont{G.}~\bibnamefont{Sarma}}, \bibinfo{journal}{J.\
  Phys.\ Chem.\ Solids} \textbf{\bibinfo{volume}{24}}, \bibinfo{pages}{1029}
  (\bibinfo{year}{1963}).

\bibitem[{\citenamefont{Gradshteyn and Ryshik}(1965)}]{GRADRYSH}
\bibinfo{author}{\bibfnamefont{I.~S.} \bibnamefont{Gradshteyn}}
  \bibnamefont{and} \bibinfo{author}{\bibfnamefont{I.~M.}
  \bibnamefont{Ryshik}}, \emph{\bibinfo{title}{Table of Integrals Series and
  Products}} (\bibinfo{publisher}{Academic Press}, \bibinfo{address}{New York},
  \bibinfo{year}{1965}).

\bibitem[{\citenamefont{Prange and Girvin}(1990)}]{PRANSTE}
\bibinfo{author}{\bibfnamefont{R.~E.} \bibnamefont{Prange}} \bibnamefont{and}
  \bibinfo{author}{\bibfnamefont{S.~M.} \bibnamefont{Girvin}},
  \emph{\bibinfo{title}{The Quantum Hall Effect}}
  (\bibinfo{publisher}{Springer}, \bibinfo{address}{New York},
  \bibinfo{year}{1990}).

\bibitem[{\citenamefont{Eilenberger}(1967)}]{EILE2}
\bibinfo{author}{\bibfnamefont{G.}~\bibnamefont{Eilenberger}},
  \bibinfo{journal}{Phys.\ Rev.} \textbf{\bibinfo{volume}{153}},
  \bibinfo{pages}{584} (\bibinfo{year}{1967}).

\bibitem[{\citenamefont{Rammer and W.Pesch}(1989)}]{RAMPESCH}
\bibinfo{author}{\bibfnamefont{J.}~\bibnamefont{Rammer}} \bibnamefont{and}
  \bibinfo{author}{\bibnamefont{W.Pesch}}, \bibinfo{journal}{J.\ Low\ Temp.\
  Phys.} \textbf{\bibinfo{volume}{77}}, \bibinfo{pages}{235}
  (\bibinfo{year}{1989}).

\bibitem[{\citenamefont{Abrikosov}(1957)}]{ABRI1}
\bibinfo{author}{\bibfnamefont{A.~A.} \bibnamefont{Abrikosov}},
  \bibinfo{journal}{Sov.\ Phys.\ JETP} \textbf{\bibinfo{volume}{5}},
  \bibinfo{pages}{1174} (\bibinfo{year}{1957}).

\bibitem[{\citenamefont{Rieck et~al.}(1991)\citenamefont{Rieck, Scharnberg, and
  Schopohl}}]{RIESCHSCH}
\bibinfo{author}{\bibfnamefont{C.~T.} \bibnamefont{Rieck}},
  \bibinfo{author}{\bibfnamefont{K.}~\bibnamefont{Scharnberg}},
  \bibnamefont{and} \bibinfo{author}{\bibfnamefont{N.}~\bibnamefont{Schopohl}},
  \bibinfo{journal}{J.\ Low\ Temp.\ Phys.} \textbf{\bibinfo{volume}{84}},
  \bibinfo{pages}{381} (\bibinfo{year}{1991}).

\bibitem[{\citenamefont{Klein}({\natexlab{a}})}]{KLEINUNP}
\bibinfo{author}{\bibfnamefont{U.}~\bibnamefont{Klein}},
  \bibinfo{note}{unpublished}.

\bibitem[{\citenamefont{Eilenberger}(1964)}]{EILE3}
\bibinfo{author}{\bibfnamefont{G.}~\bibnamefont{Eilenberger}},
  \bibinfo{journal}{Z.\ Physik} \textbf{\bibinfo{volume}{180}},
  \bibinfo{pages}{32} (\bibinfo{year}{1964}).

\bibitem[{\citenamefont{Klein}(1987)}]{KLEIN1}
\bibinfo{author}{\bibfnamefont{U.}~\bibnamefont{Klein}}, \bibinfo{journal}{J.\
  Low\ Temp.\ Phys.} \textbf{\bibinfo{volume}{69}}, \bibinfo{pages}{1}
  (\bibinfo{year}{1987}).

\bibitem[{\citenamefont{Klein}(1989)}]{KLEIN2}
\bibinfo{author}{\bibfnamefont{U.}~\bibnamefont{Klein}},
  \bibinfo{journal}{Phys.\ Rev.} \textbf{\bibinfo{volume}{B40}},
  \bibinfo{pages}{6601} (\bibinfo{year}{1989}).

\bibitem[{\citenamefont{Helfand and Werthamer}(1966)}]{HELFWERT}
\bibinfo{author}{\bibfnamefont{E.}~\bibnamefont{Helfand}} \bibnamefont{and}
  \bibinfo{author}{\bibfnamefont{N.~R.} \bibnamefont{Werthamer}},
  \bibinfo{journal}{Phys.\ Rev.} \textbf{\bibinfo{volume}{147}},
  \bibinfo{pages}{288} (\bibinfo{year}{1966}).

\bibitem[{\citenamefont{Saint-James et~al.}(1969)\citenamefont{Saint-James,
  Sarma, and Thomas}}]{SAJATH}
\bibinfo{author}{\bibfnamefont{D.}~\bibnamefont{Saint-James}},
  \bibinfo{author}{\bibfnamefont{G.}~\bibnamefont{Sarma}}, \bibnamefont{and}
  \bibinfo{author}{\bibfnamefont{E.~J.} \bibnamefont{Thomas}},
  \emph{\bibinfo{title}{Type II Superconductivity}}
  (\bibinfo{publisher}{Pergamon Press}, \bibinfo{address}{Oxford},
  \bibinfo{year}{1969}).

\bibitem[{\citenamefont{Misko et~al.}(2003)\citenamefont{Misko, Fomin,
  Devreese, and Moshchalkov}}]{MFDM}
\bibinfo{author}{\bibfnamefont{V.~R.} \bibnamefont{Misko}},
  \bibinfo{author}{\bibfnamefont{V.~M.} \bibnamefont{Fomin}},
  \bibinfo{author}{\bibfnamefont{J.~T.} \bibnamefont{Devreese}},
  \bibnamefont{and} \bibinfo{author}{\bibfnamefont{V.~V.}
  \bibnamefont{Moshchalkov}}, \bibinfo{journal}{Phys.\ Rev.\ Lett}
  \textbf{\bibinfo{volume}{90}}, \bibinfo{pages}{147003}
  (\bibinfo{year}{2003}).

\bibitem[{\citenamefont{Klein}({\natexlab{b}})}]{KLTBP}
\bibinfo{author}{\bibfnamefont{U.}~\bibnamefont{Klein}}, \bibinfo{note}{to be
  published}.

\bibitem[{\citenamefont{Klein}(2003)}]{KLEIN4}
\bibinfo{author}{\bibfnamefont{U.}~\bibnamefont{Klein}},
  \bibinfo{journal}{Physica C} \textbf{\bibinfo{volume}{388-389}},
  \bibinfo{pages}{651} (\bibinfo{year}{2003}).

\bibitem[{\citenamefont{Buzdin and Kulic}(1984)}]{BUZKUL}
\bibinfo{author}{\bibfnamefont{A.~I.} \bibnamefont{Buzdin}} \bibnamefont{and}
  \bibinfo{author}{\bibfnamefont{M.~L.} \bibnamefont{Kulic}},
  \bibinfo{journal}{J.\ Low\ Temp.\ Phys.} \textbf{\bibinfo{volume}{54}},
  \bibinfo{pages}{203} (\bibinfo{year}{1984}).

\bibitem[{\citenamefont{Mora and Combescot}(2003)}]{MORCOM}
\bibinfo{author}{\bibfnamefont{C.}~\bibnamefont{Mora}} \bibnamefont{and}
  \bibinfo{author}{\bibfnamefont{R.}~\bibnamefont{Combescot}}
  (\bibinfo{year}{2003}), \eprint{cond-mat/0306575}.

\bibitem[{\citenamefont{Casalbuoni and Nardulli}()}]{CASNAR}
\bibinfo{author}{\bibfnamefont{R.}~\bibnamefont{Casalbuoni}} \bibnamefont{and}
  \bibinfo{author}{\bibfnamefont{G.}~\bibnamefont{Nardulli}}, \bibinfo{note}{to
  be published in Rev.Mod.Phys.}

\bibitem[{\citenamefont{Shimahara}(1998)}]{HIR4}
\bibinfo{author}{\bibfnamefont{H.}~\bibnamefont{Shimahara}},
  \bibinfo{journal}{J.\ Phys.\ Soc.\ Jpn.} \textbf{\bibinfo{volume}{67}},
  \bibinfo{pages}{736} (\bibinfo{year}{1998}).

\bibitem[{\citenamefont{Nam et~al.}(1999)\citenamefont{Nam, Symington,
  Singleton, Blundell, Ardavan, Perenboom, Kurmoo, and Day}}]{NSSBAP}
\bibinfo{author}{\bibfnamefont{M.~S.} \bibnamefont{Nam}},
  \bibinfo{author}{\bibfnamefont{J.~A.} \bibnamefont{Symington}},
  \bibinfo{author}{\bibfnamefont{J.}~\bibnamefont{Singleton}},
  \bibinfo{author}{\bibfnamefont{S.~J.} \bibnamefont{Blundell}},
  \bibinfo{author}{\bibfnamefont{A.}~\bibnamefont{Ardavan}},
  \bibinfo{author}{\bibfnamefont{J.~A. A.~J.} \bibnamefont{Perenboom}},
  \bibinfo{author}{\bibfnamefont{M.}~\bibnamefont{Kurmoo}}, \bibnamefont{and}
  \bibinfo{author}{\bibfnamefont{P.}~\bibnamefont{Day}}, \bibinfo{journal}{J.\
  Phys.:\ Condens.\ Matter} \textbf{\bibinfo{volume}{11}},
  \bibinfo{pages}{L477} (\bibinfo{year}{1999}).

\bibitem[{\citenamefont{Singleton et~al.}(2000)\citenamefont{Singleton,
  Symington, Nam, Ardavan, Kurmoo, and Day}}]{SSNAKD}
\bibinfo{author}{\bibfnamefont{J.}~\bibnamefont{Singleton}},
  \bibinfo{author}{\bibfnamefont{J.~A.} \bibnamefont{Symington}},
  \bibinfo{author}{\bibfnamefont{M.~S.} \bibnamefont{Nam}},
  \bibinfo{author}{\bibfnamefont{A.}~\bibnamefont{Ardavan}},
  \bibinfo{author}{\bibfnamefont{M.}~\bibnamefont{Kurmoo}}, \bibnamefont{and}
  \bibinfo{author}{\bibfnamefont{P.}~\bibnamefont{Day}}, \bibinfo{journal}{J.\
  Phys.:\ Condens.\ Matter} \textbf{\bibinfo{volume}{12}},
  \bibinfo{pages}{L641} (\bibinfo{year}{2000}).

\bibitem[{\citenamefont{Zuo et~al.}(2000)\citenamefont{Zuo, Brooks,
  R.H.McKenzie, Schlueter, and Williams}}]{ZUOBMSW}
\bibinfo{author}{\bibfnamefont{F.}~\bibnamefont{Zuo}},
  \bibinfo{author}{\bibfnamefont{J.~S.} \bibnamefont{Brooks}},
  \bibinfo{author}{\bibnamefont{R.H.McKenzie}},
  \bibinfo{author}{\bibfnamefont{J.~A.} \bibnamefont{Schlueter}},
  \bibnamefont{and} \bibinfo{author}{\bibfnamefont{J.~M.}
  \bibnamefont{Williams}}, \bibinfo{journal}{Phys.\ Rev.}
  \textbf{\bibinfo{volume}{B61}}, \bibinfo{pages}{750} (\bibinfo{year}{2000}).

\bibitem[{\citenamefont{Shimojo et~al.}(2002)\citenamefont{Shimojo, Ishiguro,
  Yamochi, and Saito}}]{SHISYASA}
\bibinfo{author}{\bibfnamefont{Y.}~\bibnamefont{Shimojo}},
  \bibinfo{author}{\bibfnamefont{T.}~\bibnamefont{Ishiguro}},
  \bibinfo{author}{\bibfnamefont{H.}~\bibnamefont{Yamochi}}, \bibnamefont{and}
  \bibinfo{author}{\bibfnamefont{G.}~\bibnamefont{Saito}},
  \bibinfo{journal}{J.\ Phys.\ Soc.\ Jpn.} \textbf{\bibinfo{volume}{71}},
  \bibinfo{pages}{1716} (\bibinfo{year}{2002}).

\bibitem[{\citenamefont{Manalo and Klein}(2000)}]{MANKL}
\bibinfo{author}{\bibfnamefont{S.}~\bibnamefont{Manalo}} \bibnamefont{and}
  \bibinfo{author}{\bibfnamefont{U.}~\bibnamefont{Klein}},
  \bibinfo{journal}{J.\ Phys.:\ Condens.\ Matter}
  \textbf{\bibinfo{volume}{12}}, \bibinfo{pages}{L471} (\bibinfo{year}{2000}).

\bibitem[{\citenamefont{Delrieu}(1972)}]{DELRIEU}
\bibinfo{author}{\bibfnamefont{J.~M.} \bibnamefont{Delrieu}},
  \bibinfo{journal}{J.\ Low\ Temp.\ Phys.} \textbf{\bibinfo{volume}{6}},
  \bibinfo{pages}{197} (\bibinfo{year}{1972}).

\bibitem[{\citenamefont{Kramer}(1973)}]{KRAZPH}
\bibinfo{author}{\bibfnamefont{L.}~\bibnamefont{Kramer}}, \bibinfo{journal}{Z.\
  Physik} \textbf{\bibinfo{volume}{258}}, \bibinfo{pages}{367}
  (\bibinfo{year}{1973}).

\bibitem[{\citenamefont{Walker}(1988)}]{WALKER}
\bibinfo{author}{\bibfnamefont{P.~L.} \bibnamefont{Walker}}, in
  \emph{\bibinfo{booktitle}{Mathematical analysis and its applications (Kuwait,
  1985)}} (\bibinfo{publisher}{Pergamon}, \bibinfo{year}{1988}),
  vol.~\bibinfo{volume}{3} of \emph{\bibinfo{series}{KFAS Proc.Ser.}}, pp.
  \bibinfo{pages}{199--210}.

\bibitem[{\citenamefont{Buchholtz}(1969)}]{BUCHHOLTZ}
\bibinfo{author}{\bibfnamefont{H.}~\bibnamefont{Buchholtz}},
  \emph{\bibinfo{title}{The Confluent Hypergeometric Function}}
  (\bibinfo{publisher}{Springer}, \bibinfo{address}{Berlin},
  \bibinfo{year}{1969}).

\end{thebibliography}

\end{document}